# Deep Learning Methods for Colloidal Silver Nanoparticle Concentration and Size Distribution Determination from UV-Vis Extinction Spectra

Tomas Klinavičius[a*], Nadzeya Khinevich[a], Asta Tamulevičienė[a,b], Loic Vidal[c], Sigitas Tamulevičius[a,b], and Tomas Tamulevičius[a,b*]

[a]Institute of Materials Science of Kaunas University of Technology, K. Baršausko St. 59, LT-51423, Kaunas, Lithuania

[b]Department of Physics, Kaunas University of Technology, Studentų St. 50, LT-51368, Kaunas, Lithuania

[c]Institut de Science des Matériaux de Mulhouse IS2M UMR 7361, 15 rue Jean Starcky, F 68100 Mulhouse, France

*Corresponding authors Tel: +370 (37) 313432, T. Klinavičius: tomas.klinavicius@ktu.lt; T. Tamulevičius: tomas.tamulevicius@ktu.lt

## Abstract

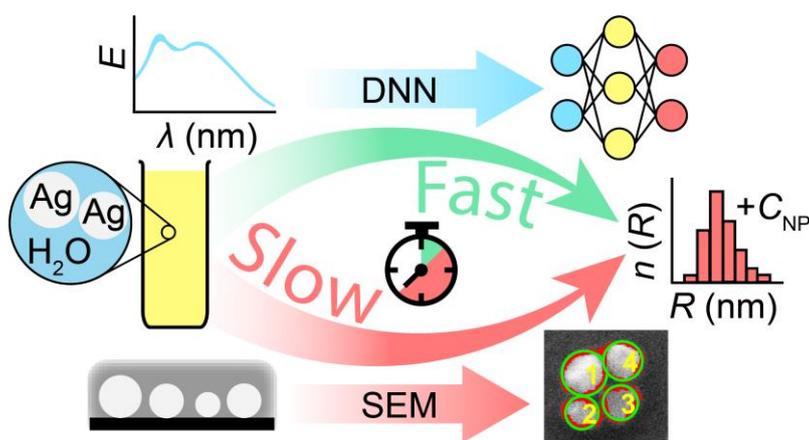

Electron microscopy, while reliable, is an expensive, slow, and inefficient technique for thorough size distribution characterization of both mono- and polydisperse colloidal nanoparticles. If rapid *in-situ* characterization of colloid samples is to be achieved, a different approach, based on fast, widely accessible, and inexpensive optical measurements such as UV-Vis spectroscopy in combination with spectra interpretation related to Mie scattering theory, is needed. In this article, we present a tandem deep neural network (DNN) for the size distribution and concentration prediction of close to spherical silver colloidal nanoparticle batches synthesized via the seeded growth method. The first DNN identified the dipole component of the localized surface plasmon resonance and the second one determined the size distribution from the isolated spectral component. The training data was engineered to be bias-free and generated numerically. High prediction accuracy with root mean square percentage error of mean size down to 1.2% was achieved, spanning the entire prediction range from 1 nm up to 150 nm in radius, suggesting the possible extension limits of the effective medium theory used for simulating the spectra. The DNN-predicted nanoparticle concentrations also were very close to the ones expected based on synthesis precursor contents as well as measured by atomic absorption spectroscopy.

**Keywords:** nanoparticles, plasmonics, silver, size distribution, deep neural networks, effective medium theory, Mie theory

## Introduction

Virtually any pure solid-state material, such as metals, semiconductors, ceramics, and others[1] or its combination can exist in the nanoparticle (NP) form[2]. NP synthesis methods can be divided into two broad

categories – physical and chemical methods[3]. Physical methods use physical interactions, such as mechanical[4] or thermal[5], to produce NPs from the bulk material. Physical synthesis methods also include laser ablation of solids in a liquid[6] and sputtering[7]. On the other hand, chemical methods use chemical reactions, such as chemical reduction[8,9] or deposition on a surface from vapor[10]. Regardless of the synthesis method, NPs always have a certain size distribution, typically a log-normal one[11,12].

Unlike microparticles, NPs exhibit size-dependent properties, which are different from those of their bulk counterparts[13]. Localized Surface Plasmon Resonance (LSPR) is a plasmonic phenomenon[14] that can be optically excited in metallic and semiconductor NPs[15]. The LSPR is observed in the extinction ($E(\lambda)$) spectrum of the NPs where multiple peaks, which correspond to different resonance modes, can manifest[16]. As a phenomenon, LSPR is highly sensitive to the refractive index of the ambient medium, making it suitable for biomolecule[17] and virus[18] detection applications. This and other unique properties of NPs make them exceptionally useful in biomedicine for the detection of antigens[19], biological warfare agents[20], and gene delivery[21]. They also found environmental and energy-related applications for cleaning up pollutants in the water[22] and conversion of $CO_2$ to carbon-based fuels[23], along with water splitting for the generation of green hydrogen[24], and enhancing solar energy harvesting[25,26]. The NP size distribution is one of the most critical parameters for their catalytic efficacy [27] as well as other cellular interactions[28], especially in cancer detection[29], and treatment[30] where knowing the NP size distribution is essential. Due to such wide application and the importance of size, simple yet rapid characterization approaches of the NP size distribution information are of special interest in an industrial setting[31].

Estimating the characteristic size dimension of NPs experimentally is generally a difficult and time-consuming task. One highly popular method is direct imaging of drop-casted NPs by scanning electron microscopy (SEM) or transmission electron microscopy (TEM)[32], as both methods measure the true geometric (Feret[33]) size of the NPs. Producing statistically reliable NP distributions by these methods requires multiple micrographs of the NPs and automatic image processing[34]. Probe-based methods such as atomic force microscopy are also used to measure the size distribution of surface-deposited NPs, but the measured sizes are inherently non-geometric due to convolution between the probe tip and the NPs[35], and are not the true NP sizes. Other size measurement techniques are optical, such as dynamic light scattering (DLS)[36]. DLS is used for measuring light scattered by NPs in a liquid medium and returns the hydrodynamic[37] size which is increased by any coating of stabilizing agents present on the NPs, thus making the method unreliable[38]. Small Angle X-ray Scattering (SAXS) is another method that uses scattered electromagnetic radiation to estimate the radius of gyration[39] of NPs, but NP sizes measured by the technique are limited to about 70 nm in diameter[40]. Many other methods of determining the size of NPs also exist[40–43], each of them with their limitations, such as price and time requirements. UV-Vis $E(\lambda)$ spectra also can be used to characterize the size distribution of liquid-suspended NPs[44] but are not straightforward to interpret[45].

The spherical NP $E(\lambda)$ spectra can be computed for a broad range of sizes either analytically employing Mie theory[36] or numerical methods[14]. Approaches based on effective media have also been developed[46]. Due to the "spectrum to distribution" problem having multiple degrees of freedom, an optimization-based approach[47] might fail to provide an optimal distribution[48]. Therefore, a more reliable approach for elucidating the size distributions from $E(\lambda)$ spectra of NPs is required.

NP concentration is another property of interest, especially for those who develop new wet-chemistry colloidal NP synthesis methods[49,50]. NP seed solutions concentration determination is paramount for computing volumes and chemical precursor concentrations needed to produce specific-size NPs via the seeded growth method[51]. Rapid *in-situ* determination of concentration using spectroscopic data is of particular interest[52].

In recent years, machine learning (ML) based data analysis tools have exploded in popularity[53]. As a data-driven approach, an ML algorithm can learn the underlying patterns in the data and approximate them effectively, circumventing computational or analytical difficulties[54,55]. Recently, ML has emerged as a popular method for the characterization of particle size distributions[56,57]. Some authors use pre-processing of $E(\lambda)$ spectra in order to increase the accuracy of ML algorithms[57] while others use the raw spectra[44]. Using computationally derived data to train ML algorithms has also drawn attention[56,58]. ML tools such as DNNs have been proven to be exceptionally powerful[59]. Due to the wide functionality of DNNs[60], they have found many scientific applications in biomedicine[61], materials science[62], detection of chemical materials[63], and many more fields, including design of NP colloids[64], and assisting in their synthesis[65,66]. Early attempts at using multilayer feedforward DNNs to analyze particle sizes focused on the microparticle size range[67–70]. In later years, both DNNs and other ML techniques were used to investigate size distributions of particles in the nanoscale[71], including more complex shapes such as rod-like NP[56,72], but their success was limited at best. Essentially, determining the size distribution from an $E(\lambda)$ spectrum is an inverse modelling problem. A tandem deep neural network (DNN) approach is used quite often for solving inverse problems, for example in photonic design[73–76].

In this work, we present a tandem DNN architecture for determining the radius distribution of seeded growth synthesized silver (Ag) colloidal NPs as well as their concentration in the colloid from UV-Vis $E(\lambda)$ spectra. The DNNs were trained using computationally generated data composed of NP distributions, concentration values, and $E(\lambda)$ spectra. The data was engineered to be as bias-free as possible by thoroughly examining the influence of data parameters on the characteristic aspects of the data and selecting said parameters to produce unbiased data instances for the training data. The tandem DNN was tested with experimental wet-chemistry synthesized NP data, providing an excellent match between predictions and empirical results. The first stage of the tandem DNN processed the spectra by first isolating the dipolar peak of the spectrum and the second stage performing the prediction. Using the selected tandem architecture, we have demonstrated that accurate predictions of NP size can be made in a broad range – up to 150 nm in radius, more than other authors – by using only dipole, instead of dipole and quadrupole, features. What is more, our approach allowed to estimate the concentration of both the total mass of Ag and the Ag NPs concentration in the solution.

# Methods

## Experimental Methods

**Synthesis of NPs**

Ag NPs were synthesized using the seeded growth method[8] (**Figure 1a**). The size and size dispersion of Ag NP seeds were controlled by varying the concentrations of trisodium citrate (TC, $C_{TC}$), tannic acid (TA, $C_{TA}$), and silver nitrate (SN, $C_{SN}$) (all from Sigma Aldrich) in the aqueous solution. Ag NPs were grown from seeds (**Figure 1b**) and enlarged in sequential synthesis steps (**Figure 1c**). More details on both seeds and NP synthesis are provided in **Supplementary Section S1.1**. Six different NP colloid batches ("A"-"F") were produced as summarized in **Table S1**. Batches "A" – "E" were synthesized using varying concentrations of TA in the seed solution to induce NP growth, resulting in a greater variation in their size distribution. On the other hand, NPs from batch "F" were synthesized with a reduced TA concentration in the seed solution aiming for a decrease in the concentration of components during the growth process for approaching consistent NP growth and maintaining a constant standard deviation in radiuses. The production of a wide variation in NP sizes, along with varying size distributions, served as a means to validate the proper functionality of the DNNs.

**Characterization of NPs**

*UV-Vis spectroscopy of Ag colloids*

A 1.4 nm spectral resolution optical fiber spectrometer "AvaSpec-2048" (Avantes) and light source "AvaLight-DHc" (Avantes) were used for measuring the $E(\lambda)$ spectra of colloid solutions in 1 mm optical path length quartz cuvette. Initially, colloids were measured with their original particle concentration. All $E(\lambda)$ measurement results were normalized to the $E(\lambda)$ of the same cuvette filled with water. All colloids were sonicated prior to UV-Vis measurement in order to break up possible agglomerates.

*Determination of NP size distribution*

Before imaging, the NP colloids of their original concentration were sonicated in order to break up possible agglomerates. Before the drop-casting of NPs for imaging, the colloid solution was processed in order to remove the TC surfactant. 100 μL of the original solution was centrifugated at 7000 rpm for 7 min. After centrifugation, the solution-supernate was replaced with 100 μl of ethanol, the mixture was sonicated and centrifugated at the same parameters. This procedure was repeated twice and after the second centrifugation in ethanol, the sediment was extracted and placed on a piece of silicon wafer. Each NP sample was prepared for scanning electron microscope (SEM) imaging by drop-casting the corresponding colloid solution on a silicon substrate. Field emission gun SEM "Quanta 200 FEG" (FEI) was used. Multiple micrographs of NPs from every sample were made in order to get a sufficient amount of NP images for a reliable effective NP radius statistical analysis. Effective Ag NPs radius distribution was derived from SEM micrographs employing an automated custom micrograph analysis and statistics processing script that was implemented in MATLAB (MathWorks), similar to[34]. The principle of SEM micrograph pre- and post-processing is showcased in **Figure 1d** – the red outline shows the actual contour of the NPs while the green outline is the circle with the resulting effective NP radius. The steps of the automated analysis are depicted in **Figure S1** and described in more detail in **Supplementary Section S1.2**. The hydrodynamic particle radius and its distribution were also confirmed by DLS employing a "Zeta Sizer Ultra" particle size analyzer (Malvern Panalytical). Details of the measurement can be found in **Supplementary Section S1.3**.

*Determination of NP shape and Crystallinity*

A high-resolution transmission electron microscope (TEM) "ARM-200F" (JEOL) was used for imaging Ag seeds and NPs. NPs were prepared for imaging by casting them on a TEM wire grid and then drying them at room temperature. Due to the higher magnification of the TEM when compared to SEM, TEM micrographs were used for NP shape and qualitative crystallinity inspection purposes. The crystallinity of several selected samples was also evaluated by X-ray diffraction (XRD) using an X-ray diffractometer D8 Discover" (Bruker AXS GmbH). Measurement details can be found in **Supplementary Section S1.4**.

*Concentration measurement of colloids*

Atomic absorption spectrometer "AAnalyst 400" (PerkinElmer) was used to measure the mass concentration (mg/L) of Ag in the Ag NP colloids. Colloid samples were specially prepared before measurement – Ag NPs were dissolved in Aqua Regia mixed in a ratio of 2 to 1.

## Numerical Methods

### Computing Hardware and Software

A desktop computer with a processor "Intel(R) Core (TM) i5-9500 CPU @ 3.00GHz" and 16 GB of RAM running "MATLAB 2021a" (MathWorks) on "Windows 10 Enterprise" (Microsoft) was used to perform all computations. Training of the DNNs as well as their hyperparameter optimization was performed on the same computer, using a graphical processing unit (GPU) "GeForce GTX 1050 Ti" (NVIDIA).

### Mie Theory for Calculation of Full Extinction Spectra

Mie theory, as described in[36], was used to compute the $E(\lambda)$ of spherical NPs. The spectra $E(\lambda)$ contained peaks corresponding to all resonances depending on NP size, not just the dipolar ones. Such spectra in this work are referred to as *full* spectra (**Figure 1f** blue spectra). The NP distributions used to originate *full* spectra were the same as those used for *dipole* spectra. In Mie theory, the amplitude of $E(\lambda)$ is defined by the number of NPs ($N$) interacting with the light wave and the thickness of the interaction region (constant in our case). The Palik permittivity was used to define the Ag NPs[77] while the permittivity of water was taken from[78].

### Transfer Matrix Method Using Effective Medium Theory for Calculation of Dipolar Extinction Spectra

The Transfer Matrix Method (TMM)[79] based on scattering matrices[80] was used to compute $E(\lambda)$ containing only dipolar resonances corresponding to $E(\lambda)$ peaks of spherical NPs. In this work, such spectra are referred to as *dipole* spectra (**Figure 1f** green spectra). Since TMM is only able to perform computations on homogeneous slabs of material, a modified Maxwell – Garnett – Mie (MMGM) effective medium theory (EMT)[46] was used to homogenize the colloidal solution of randomly distributed subwavelength-sized NPs into a homogeneous medium (**Figure 1f**). The same permittivity dispersions were used as in Section 2.2.2. Using an EMT allows to avoid a generally computationally intensive and time-consuming 3D numerical simulation.

This was extremely useful for generating training and validation data in vast quantities. The effective medium approximation is valid as long as the NPs are of sufficiently subwavelength size[81]. The effective medium model and its validity for larger NPs are described in more detail in **Supplementary Section S1.5**.

In MMGM, concentration is accounted for as a volume-filling fraction (*F*). The concentration of NPs in the colloid directly impacts the amplitude of the modelled spectra, with a higher concentration corresponding to a higher amplitude as explained in detail in **Supplementary Section S2.5**.

**Engineering of Training and Validation Data**

Using Mie theory for *full* spectra and TMM with MMGM medium for *diple* spectra, NP distributions, and corresponding $E(\lambda)$ spectra were synthetically generated for both training and validation of DNNs. Assuming the log-normal shape of NP size distributions, the mode ($M_L$) and full width at half maximum (FWHM$_L$) were selected as the parameters defining the distribution. Due to lack of a parametrization of log-normal distributions with these parameters, they were subsequently transformed into the canonical parameters, mean $\mu_L$ and standard deviation $\sigma_L$, which were used to generate the distributions. The derived equations of the transformation (eq. S4) – (eq. S5) are provided in **Supplementary Section S1.6** while the ranges of all data parameters are found in **Table S2**.

Two sets of computationally derived training and validation data were generated – one for "DipoleNet" and one for "ColloidNet". For "DipoleNet", data consisted of pairs of *full* spectra (inputs, **Figure 1f** blue spectra) and corresponding *dipole* spectra (outputs, **Figure 1f** green spectra). For "ColloidNet", data pairs were composed of *dipole* spectra and corresponding NP distributions along with their *F* values (**Figure 1f** red distribution). The size of each training dataset was 626 359 pairs, while the size of each validation dataset was 156 590 pairs – 1/4$^{th}$ the size of the training data.

Testing data for "DipoleNet" used experimentally obtained *full* $E(\lambda)$ spectra for inputs. For the output, the best *dipole* spectrum fits, computed with TMM using MMGM medium, for the *dipolar* resonance peak of *full* spectra were used. Testing data for "ColloidNet" consisted of *dipole* spectra obtained from "DipoleNet" for the inputs and experimentally obtained distributions of effective NP radius and NP colloid concentration measurements for the outputs. Each testing dataset consisted of 45 corresponding pairs.

**The Architecture of Neural Networks, Hyperparameter Optimization and Network Training**

DNN "DipoleNet" is constructed as a multilayer feedforward neural network and is depicted in **Figure 1e**. It is composed of an input layer, internal blocks of layers and an output layer. A fully connected layer followed by a Leaky ReLU activation function constitutes a single block. The Leaky ReLU function was chosen to avoid the vanishing gradient problem[82]. Finally, a regression layer was used as the output layer.

Similarly, "ColloidNet", depicted in **Figure 1e**, was also composed of an input layer, repeating layer blocks and an output layer. This time, a single block was composed of a fully connected layer, a normalization layer, a Leaky ReLU layer, and a dropout layer. Descriptions of all layers are available in[83].

To determine the best hyperparameters[84], hyperparameter optimization (HPO) using Bayesian optimization[85] was performed. The numbered workflow of HPO for both DNNs is depicted in **Figure 1g**. For DNN training during HPO, 1/10$^{th}$ of the training and validation datasets are randomly extracted into reduced training and validation datasets. Validation RMSE was chosen as the cost function to minimize. The reduced datasets were reshuffled once every epoch to ensure maximum data variation during training. It is assumed that hyperparameters which assure the best network performance (lowest validation root-mean-square error, RMSE) will also provide the best network performance when training and validating the DNN with full-sized datasets. It was empirically determined that the root mean square propagation training algorithm[86] works best for training both DNNs. After HPO, optimal hyperparameters were used to train and monitor the DNNs with their respective full-sized training and validation datasets. The same training parameters that were used during HPO were used for the final training process. Ranges of hyperparameters, along with optimal values and other parameters of both DNNs are listed in **Table S3.** The use of training and validation data during training is described in **Supplementary Section S1.7**.

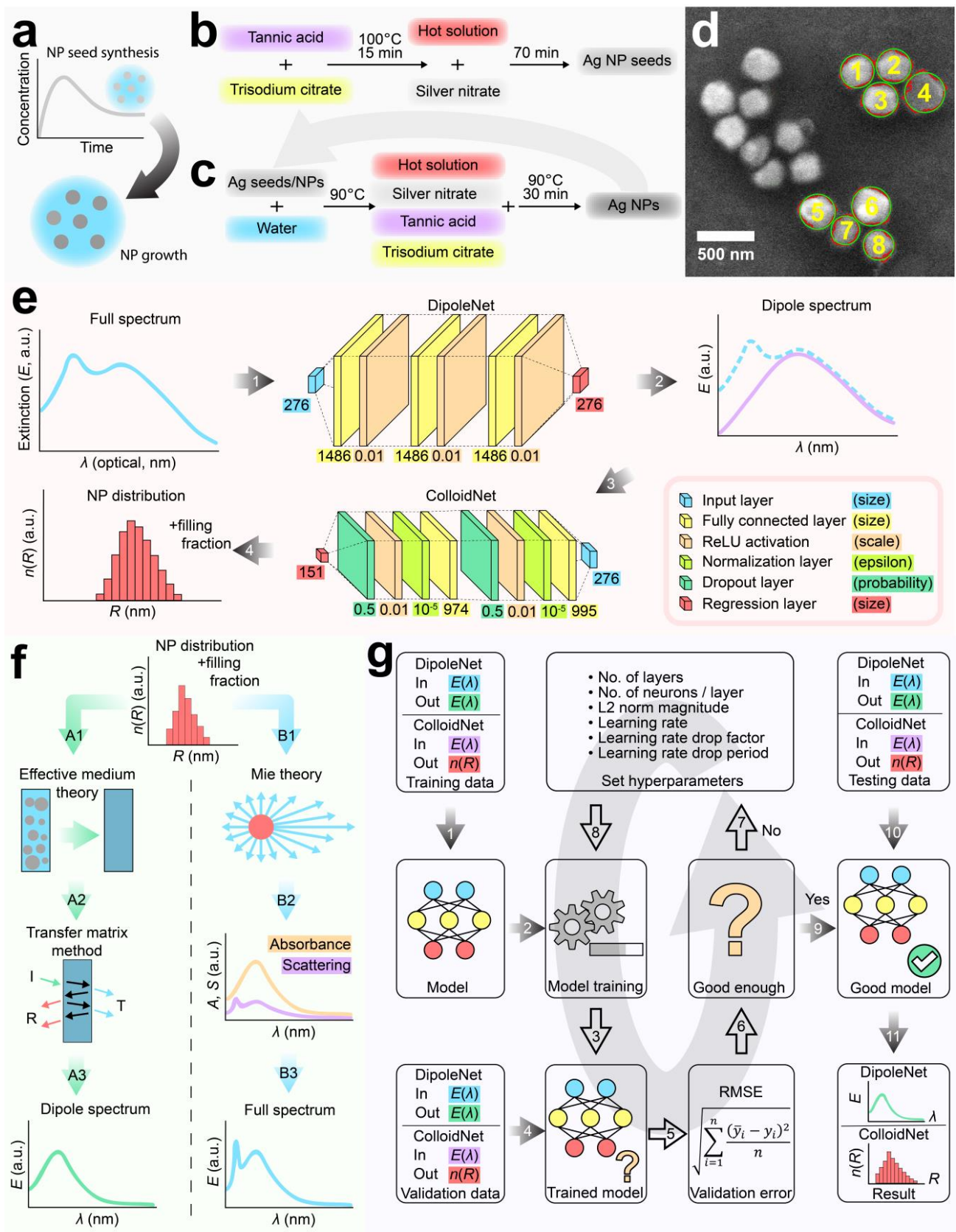

**Figure 1.** Experimental and computational methods used for Ag NP colloid $E(\lambda)$ spectra analysis for obtaining the effective NP radius size distribution ($n(R)$). (a) Graphical representation of Ag NP seed synthesis and subsequent growth. (b) The chemical reaction for the Ag NP seeds colloid synthesis. (c) The reaction for Ag NPs synthesis. (d) Determination of the empirical NP effective radius distribution from SEM micrographs, and overlaid graphics indicates the automated area analysis results used for extraction of the effective radius distribution of numbered NPs. (e) Working principle of the proposed tandem DNN system for determining NP concentration and radius distribution in a colloidal solution from its $E(\lambda)$ spectra, the legend explains the layers and their parameters of the DNNs used where "DipoleNet" is responsible for subtracting the dipole peak and the "ColloidNet" for finding the radius distribution from the pretreated $E(\lambda)$ spectra, arrows indicate the order of operations (input of extinction spectrum to "DipoleNet" – "1", obtaining output from "DipoleNet" – "2", inputting output of "DipoleNet" to "ColloidNet" – "3" and obtaining output from "ColloidNet" – "4"). (f) Generation of training data for both DNNs using effective medium theory with TMM (computation of effective NP permittivity – "A1", input of effective medium to TMM – "A2" and computation using TMM – "A3") and Mie theory (computation of parameters for Mie theory using NP distribution – "B1" and performing computations of absorbance and scattering "B2" to obtain $E(\lambda)$ "B3"). (g) Workflow for hyperparameter optimization and network testing for both networks (input of training data "1" to the model "2", training the model "3", inputing validation data to a trained model "4", obtaining validation error "5", comparing to a desired performance metric "6", setting different hyperparameters "7" if performance was not satisfactory and using new hyperparameters for training "8" or, if performance is satisfactory, obtaining a trained model "9" which is tested with testing data "10", to obtain outputs (size and concentration analysis results) "11").

## Results and Discussion

### Experimental results

Overall, 45 Ag NP samples were synthesized via the seeded-growth method (**Figure 1a, b, c**)[8]. NPs were specifically synthesized to have 2 distinct kinds of distributions – narrow distributions and broad distributions – for various mean radius ranges. This distinction between different batches is clearly visible in the distributions of their statistical parameters, depicted in **Figure S2** and available in **Table S1**. From this empirical data, it was observed that radius distributions with a smaller mean NP radius are narrower, while distributions with a larger mean NP radius are broader. This directly translates into the full width at half maximum (FWHM) of the log-normal fits and is consistent with experimental data provided by other authors[8]. Meanwhile, NP radius distributions (histograms of probability density with a bin width of 1 nm) for all samples (sorted by mean NP radius) are displayed in **Figure 2e**. Mean NP radius values in the study span from 10.5 nm to 107.6 nm. On average, around 1000 (and sometimes up to *ca.* 2000), but never less than 600, NPs are observed for each NP sample, providing high reliability of the collected statistics of NP radius distributions as recommended by the literature[87]. The strong reliance on a number of NPs imaged for obtaining accurate size distributions is one of the biggest drawbacks of direct imaging methods like SEM or TEM. The SEM-measured NP size distributions were verified by TEM analysis and indicated a good overall match between the results, but due to a significantly smaller number of NP samples analyzed were not used further and are not shown here. Based on the goodness of fit being $R^2 > 0.87$ for all samples, it was confirmed that the NP radius distributions have a log-normal distribution. Such observation is consistent with the literature[11,12]. Another drawback of direct imaging techniques like SEM is that the mean size of NPs is subject to shifting depending on how the micrographs are analyzed. Finally, NP samples can require specific preparation such as surfactant stripping or sonication, both of which were used, in order to obtain the best possible results from SEM.

Because imaging-based measurement techniques (TEM, SEM) determine the size of each NP directly, the quality/noise of the underlying distribution shape and reliability of statistical parameters derived from said distribution are affected by the number of NPs that are analyzed. Indirect measurement methods such as DLS[36] produce smooth NP distributions. They confirmed the tendency of the electron microscopy results, but despite that, DLS-determined size distributions did not always absolutely overlap with electron microscopy results (**Figure S3**) and were sub-optimal for DNN validation as explained in more detail in **Supplementary Section S2.2**.

Both transmission and scanning electron micropcopies confirmed that the NPs seeds were spherical (**Figure 2a**) and became more faceted as NPs grew (**Figure 2b**) but in general preserved a symmetrical close to spherical shape, hence suitable to be described by the MMGM effective medium theory[46]. This is consistent with the literature describing the synthesis of spherical Ag NPs using various wet chemical methods[88–91]. In **Figure 2c**, the crystallite planes corresponding to the Ag (111) facet can be attributed to sub-NP-sized crystallites that constitute the polycrystalline NPs and are larger for bigger radius NPs (**Figure S4**) as discussed in more detail in **Supplementary Section S2.3**.

$E(\lambda)$ spectra following the same increasing mean NP radius order are displayed in **Figure 2d**. For narrow NP distributions with a small mean NP radius, the $E(\lambda)$ spectrum consists of a single narrow peak, corresponding to a dipole resonance[92]. As the mean NP radius increases, the peak $E(\lambda)$ wavelength shifts towards longer wavelengths, and the FWHM of the $E(\lambda)$ peak increases due to dynamic effects[46]. In general, the $E(\lambda)$ spectrum is composed of a multitude of resonances (**Figure S5 a**), but for small NPs only dipolar resonances manifest. As NP size increases, higher-order LSPR oscillations become supported – a quadrupole $E(\lambda)$ peak emerges and continues to grow in intensity, peak position, and FWHM, eventually overtaking the *dipole* peak's intensity (**Figure S6 b**). These findings are consistent with widely known predictions made with Mie theory reported by other authors, *e.g.*,[36,93]. Once higher-order modes are supported, the quasi-static approximation[16] no longer holds. At that point, scattering is the dominant factor of $E(\lambda)$ (**Figure S6 a**). However, this transition is not clear-cut, as determined in[81]. In our work, it was found that the MMGM effective medium theory is still valid even for NPs with a radius of 150 nm, as long as $F$ is increased to account for the loss of amplitude. For a more in-depth discussion about the validity of the effective medium theory used for generating synthetic $E(\lambda)$ spectra, see **Supplementary Section S2.4**.

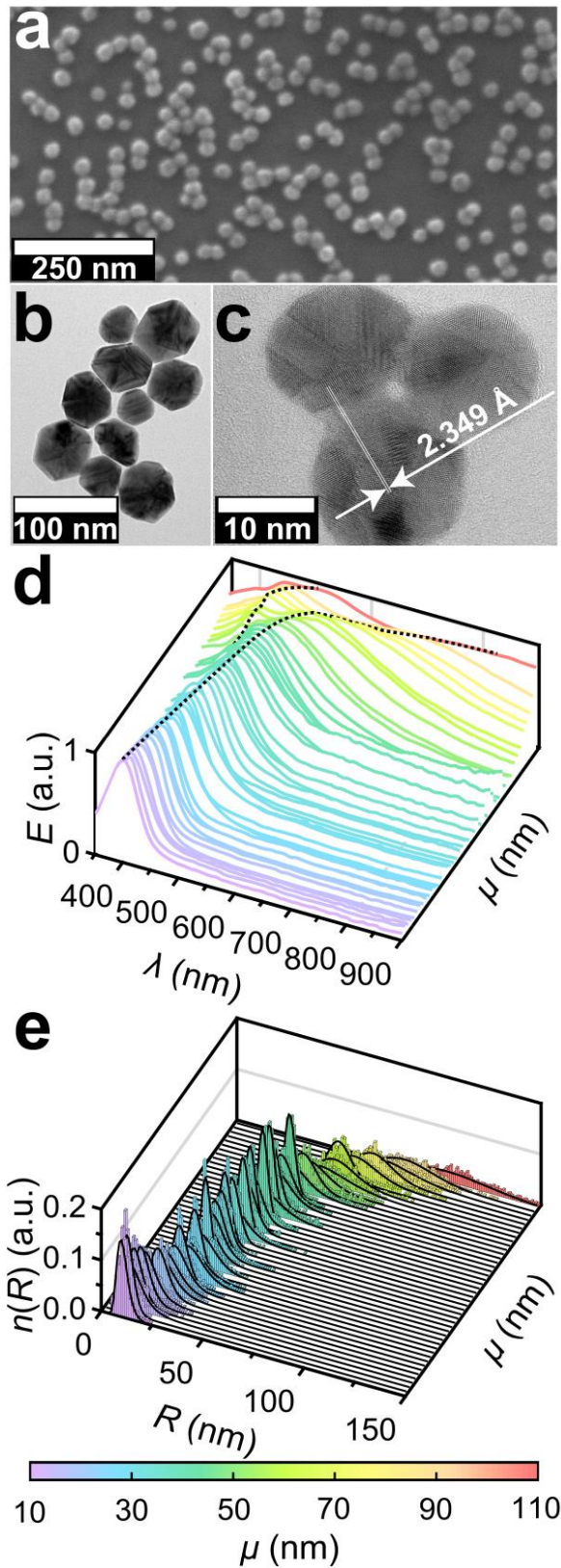

**Figure 2.** Properties of the synthesized Ag NP colloids. (a) Typical SEM micrograph of Ag NPs from a drop casted colloidal solution on a silicon substrate (sample "C5"). (b) TEM micrographs of NPs after several stages of growth (sample "F15"). (c) High-resolution TEM micrographs of NP seeds for the "F" batch, showing the crystallite boundaries of the NP seeds with indicated interatomic distance corresponding to the Ag (111) plane. (d) $E(\lambda)$ spectra, normalized to one, of all investigated colloidal solutions with increasing mean NP radius, truncated lines indicate *dipole* and quadrupole peak positions. (e) NP radius distributions corresponding to the spectra in (d) extracted from the SEM micrographs, a solid line represents the log-normal fit of experimental distribution.

## Numerical Data Results

Using equations (eq. S4) – (eq. S5) and numerically derived relationships of $F$ (eq. S6) and $N$ (eq. S8), data instances were engineered to be as bias-free as possible by selecting the ranges of the underlying data values to correspond to those of testing data and by equally representing all cases of data values in those ranges (**Table S2**). The relevant data parameter region in the ($M_L$, $FWHM_L$) space is depicted in **Figure 3c** (top), and the same region post-transformation in the ($\mu_L$, $\sigma_L$) parameter space is depicted in **Figure 3c** (bottom). Due to the logarithmic nature of the log-normal distribution parameters ($\mu_L$, $\sigma_L$), the post-transformation region is spaced unevenly compared to the linear ($M_L$, $FWHM_L$) space.

The effect of ($\mu_L$, $\sigma_L$) on $F$ in MMGM when the spectrum amplitude is maintained is highly non-linear, as depicted in **Figure 3b** and in **Figure 3d** in more detail. An equivalent relation also exists for $N$ in Mie theory. This was taken into account when generating training and validation data. It is observed that for larger NPs $F$ has to increase tremendously in order to preserve the amplitude. This is because the absorbance cross-section decreases in relation to the $E(\lambda)$ cross-section due to the onset of scattering (**Figure S6**). That is why the same value of $F$ produces a smaller amplitude of the spectrum for larger NPs as computed with TMM and MMGM – because MMGM computes absorbance and is based on the 1$^{st}$ (dipolar) electric Mie coefficient. However, this change in amplitude does not have any effect on the shape of the spectrum, as it is an extremely close match when the relevant peak of *full* and *dipole* $E(\lambda)$ spectra is brought to the same amplitude (**Figure 3a**). Due to the effect of $F$ weakening with increasing NP size, it needs to be increased in order to maintain the same spectrum amplitude. Knowing the dependence of $F$ on the $E(\lambda)$ amplitude to be directly proportional[46], this dependence is universal for any spectrum amplitude. An analogous dependence was determined for $N$, which is a parameter in Mie theory that is equivalent to $F$ in MMGM. Due to scattering being accounted for in Mie theory, the dependence of $N$ was different than that of $F$. Both of these dependencies were used to generate synthetic spectra with amplitude changing in an exact manner. This allowed for generating training and validation data with a high amount of variation and virtually no bias.

Due to the dependence depicted in **Figure 3d** being derived numerically, it could not be used as a general expression, which is crucial for generating the $F$ value for any given parameters ($\mu_L$, $\sigma_L$). Therefore, it was fitted using a 2 variable 5$^{th}$ order polynomial fit, with goodness of fit parameter $R^2$ being 0.999. For the actual application, the fit of $F$ was used to compute the value of $F$ for log-normal distributions with parameters ($\mu_L$, $\sigma_L$).

It should be noted that NPs have a certain lower size limit where they transition into clusters, properties of which are quantum in nature. This transition is usually considered to happen at *ca.* 1 nm radius[94–97] (see **Supplementary Section S1.5** for more details), hence it is the smallest radius value chosen to be predicted by the tandem DNN.

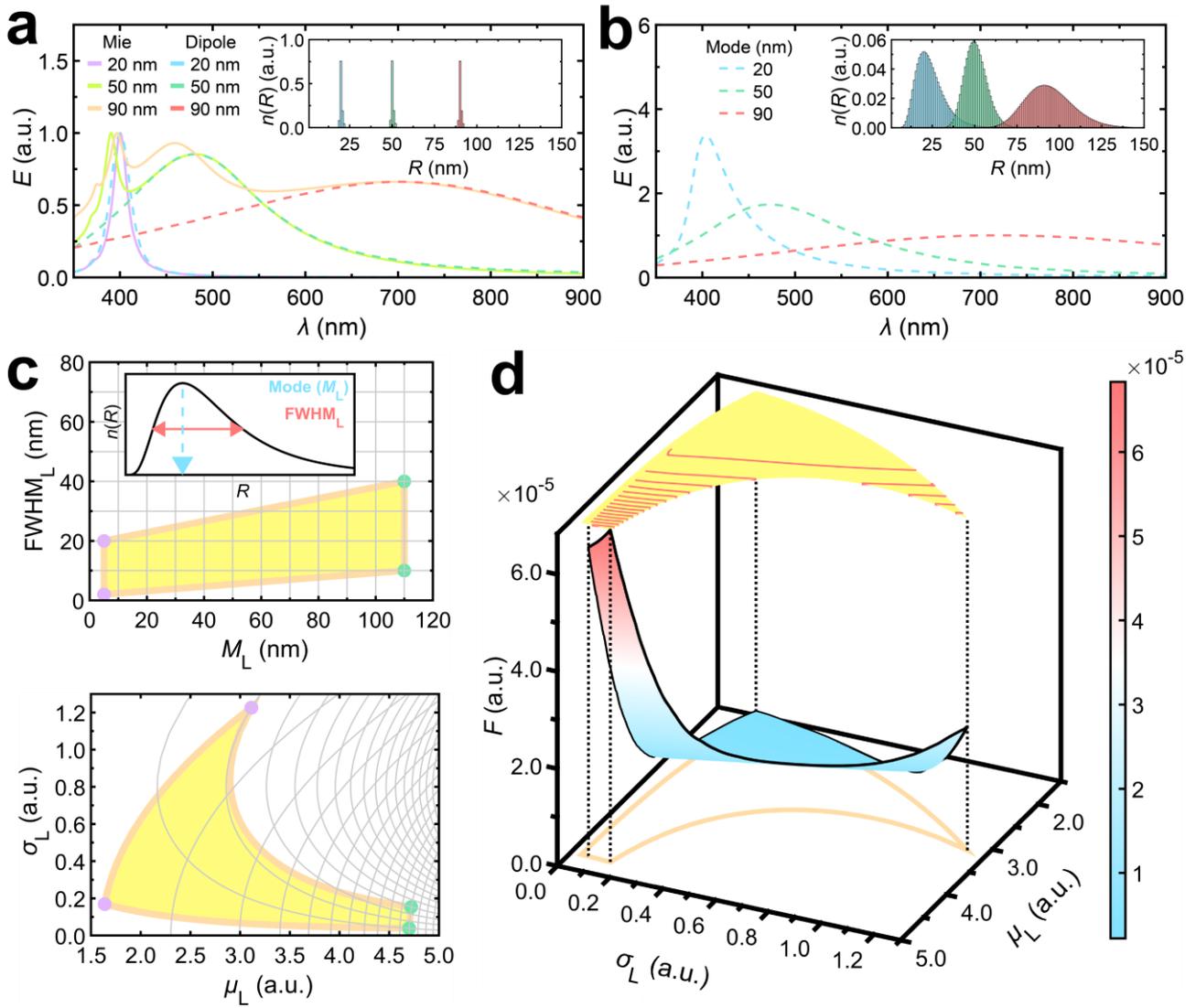

**Figure 3.** Training data for DNNs. (a) Comparison of $E(\lambda)$ *full* spectra originated with Mie theory (solid lines) and *dipole* spectra originated with TMM using MMGM medium (dashed lines), the inset depicts narrow radius distributions ($M_L$ 20, 50, and 90 nm, $FWHM_L$ 5 nm) used to compute $E(\lambda)$ spectra employing both origination methods. (b) *Dipole* $E(\lambda)$ spectra for broad radius distribution NPs with $M_L$ identical to those used in (a), depicted in the inset. Spectra were computed with TMM using MMGM medium where the volume filling fraction doubles for each subsequent distribution. (c) Log-normal distribution parameter ($M_L$ and $FWHM_L$) values in the training data before and after transformation to canonical distribution parameters $\mu_L$ and $\sigma_L$. (d) Values of filling fraction $F$ depending on canonical distribution parameters required for achieving the amplitude of *dipole* $E(\lambda)$ spectrum equal to one.

## "DipoleNet" DNN Results

The results of the "DipoleNet" tests are summarized in **Figure 4**. It is important to mention that the output of "DipoleNet" is purely fictitious as there are no real *dipole* $E(\lambda)$ spectra[98] and is only used as a stepping stone for achieving the end result – the output of "ColloidNet". During the operation of the tandem DNN, this output is further used as input for "ColloidNet".

As seen in **Figure 4a**, the $E(\lambda)$ peak which corresponds to higher-order charge oscillations is becoming more prominent with increasing NP size. The results of using "DipoleNet" on its testing data can be seen in **Figure 4b**. The residual between the *dipole* fit spectrum and the prediction of "DipoleNet" at most is about 2 orders of magnitude less than the $E(\lambda)$ amplitude, which corresponds to an absolute error of a few percent at most. This indicates that the "DipoleNet" has correctly learned to predict the characteristics of the $E(\lambda)$ peak which corresponds to the *dipole* component of the $E(\lambda)$. Furthermore, in **Figure 4c** it is seen that the root mean square percentage error (RMSPE,[99]) between the *dipole* fit and "DipoleNet" prediction is also consistently small throughout a wide range of NP radii. Not only does this further support the conclusion that "DipoleNet" has correctly learned to identify the dipole feature, but it also shows the network's reliability for NPs of various sizes. "DipoleNet" conducted each individual prediction in about 0.0031 s.

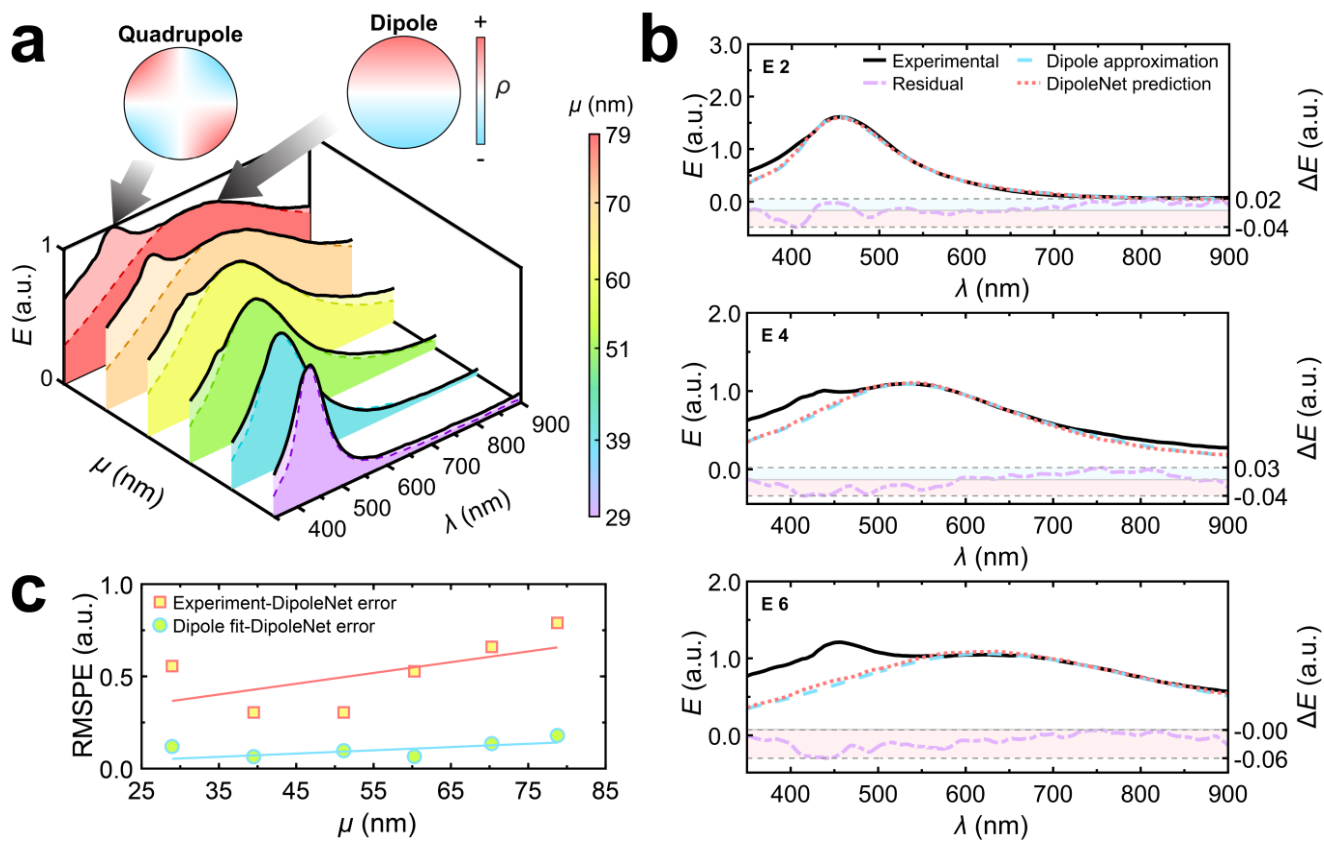

**Figure 4.** Extraction of a *dipole* component from *full* $E(\lambda)$ spectrum. (a) Examples of *full* spectra (light shade) and *dipole* spectral components (dark shade with dashed line) for the "E" batch of colloid solutions increasing in mean NP radius extracted with "DipoleNet", along with surface charge density distributions corresponding to distinct resonances. (b) The 2$^{nd}$, 4$^{th}$, and 6$^{th}$ spectra from (a) compared with *dipole* spectra predicted by "DipoleNet" and *dipole* spectra computed with effective medium theory and TMM for the best fit of the *dipole* spectral peak on the primary axis, and the difference between *dipole* spectrum and predicted spectrum of "DipoleNet" on the secondary axis. (c) RMSPE between *full* $E(\lambda)$ spectra and "DipoleNet" predicted $E(\lambda)$ spectra compared to the error between "DipoleNet" spectra and *dipole* spectra for various mean NP sizes.

## "ColloidNet" DNN Results

### Predictions of NP size

The "DipoleNet"-processed experimental $E(\lambda)$ spectra were fed as inputs to "ColloidNet" and resulted in the NP size predictions depicted in **Figure 5**. Similarly to "DipoleNet", "ColloidNet" originated its predictions in *ca.* 0.0082 s per spectra. As seen in **Figure 5a**, all distributions have a clearly defined log-normal nature[11]. By mere visual inspection, it can be seen that the shape of the predicted distributions is close to empirically determined NP radius distribution indicating that "ColloidNet" predictions are correct. The main source of mismatch between the ground truth (experimental data) and network predictions seems to be noise due to having a limited number of samples as *ca.* 1000 NP radius data points might still be not enough to produce a smooth histogram and/or possible analysis errors of the automated SEM micrograph processing algorithm. There are two possible error types in the micrograph processing algorithm – (i) over-segmentation of NPs when one NP is misunderstood as several smaller ones and (ii) non-separation where several NPs clumped together are misunderstood as one large NP. Both can arise due to using the watershed transformation[100] to separate the NPs clumped together in the micrographs.

The true – predicted plots corresponding to the three analyzed histograms in **Figure 5a** are depicted in **Figure 5b**. They confirm a good correlation between the predicted values and the ground truth. Most points follow the trend of this line while overestimates of the histogram bins are above the ground truth line and underestimates are below. These errors mostly represent the smallest values of $n(R)$ (see **Figure 5b**) that are less significant as they have little contribution to $E(\lambda)$. It can be further seen that for distributions with increasing mean NP size, the RMSPE is consistently under 50%, indicating the "ColloidNet" has a similar level of prediction accuracy for NPs over a broad size range. Additional results of Ag NP size prediction for "C" series (**Figure S8**) and for the mean size of all investigated colloids (**Figure S9**) can be found in **Supplementary Section S2.6**.

The quantitative statistical distribution parameters – the mean ($\mu$), the mode ($M$), and the standard deviation ($\sigma$) – for both experimental and predicted distributions are provided in **Figure 5a** legends for selected cases and also analyzed for the entire batch "E" in **Figures 5c**, **d**, and **e**, respectively, indicate excellent agreement between predicted and experimental values. The mean $\mu$ was predicted with the highest level of accuracy while $M$ was slightly underestimated. Haiss et al.[101], proposed a highly simplified method for determining mean NP sizes from their dipolar $E(\lambda)$ peak spectral location, but it makes no effort to account for the shape and especially the size distribution width. Meanwhile $\sigma$ had the largest deviation from the ground truth and was always overestimated. Overestimation of $\sigma$ values can be explained by the noise in network predictions. Although not visible in the histograms due to its small size, predictions made with "ColloidNet" contain low amplitude noise, which could have influenced the estimate of $\sigma$. Another possibility is that NPs much larger or smaller than the mean values were simply not observed due to an insufficient number of NP radius samples, as it would require using tens or hundreds of thousands of samples to observe them because they are just that rare. Still, value prediction is consistently accurate over a large interval, and in most cases, the parameters are within 10% RMSPE bounds or even as small as 1.2% as for $\mu$ of large NPs. The overall prediction accuracy for $\mu$ in the study was 6.1% (**Figure S9**) and was consistently better for larger NPs. This shows the versatility of the proposed tandem DNN system. K. Shiratori et al.[56] reported a similar prediction accuracy level for the nanorod sizes. This work has the advantage over[71] because here size characterization is achieved only from UV-Vis $E(\lambda)$ spectra, without requiring DLS data as input. Moreover, to the best of our knowledge, current work exceeds the maximum plasmonic NP radius prediction range of 110 nm previously reported by E. X. Tan et al.[57] by at least 40 nm. Higher prediction accuracy without sacrificing prediction range could potentially be achieved by using empirical $E(\lambda)$ spectra as training data, but this requires the synthesis of tens if not hundreds of colloids with precision NP size tuning and concentration increment[57]. Moreover, the current study was limited by the spectral range of the silicon-based detector spectrometer and deuterium-tungsten lights source while the LSPR range together with mean sizes could be potentially extended using a multiple detector UV-Vis-NIR spectrometer. Using both dipolar and quadrupolar features/spectra could potentially yield an even wider window of size characterization[57], but it would require discarding the use of the effective medium theory which brought computational benefits met in this work.

The insets of **Figure 5a** depict the *full* experimental UV-Vis $E(\lambda)$ spectra along with the DNN predicted *dipole* spectra while it is a direct computation result of the "ColloidNet" predicted parameters – NP radii distributions and their respective volume filling fractions. The shape of the *dipole* peak is mainly affected by the NP radius distribution shape while the amplitude of the peak is affected by the filling fraction, which is equivalent to the NP concentration. A good fit of the *dipole* peak is one more quality estimation metric indicator that the prediction is correct or close to it, especially in terms of concentration.

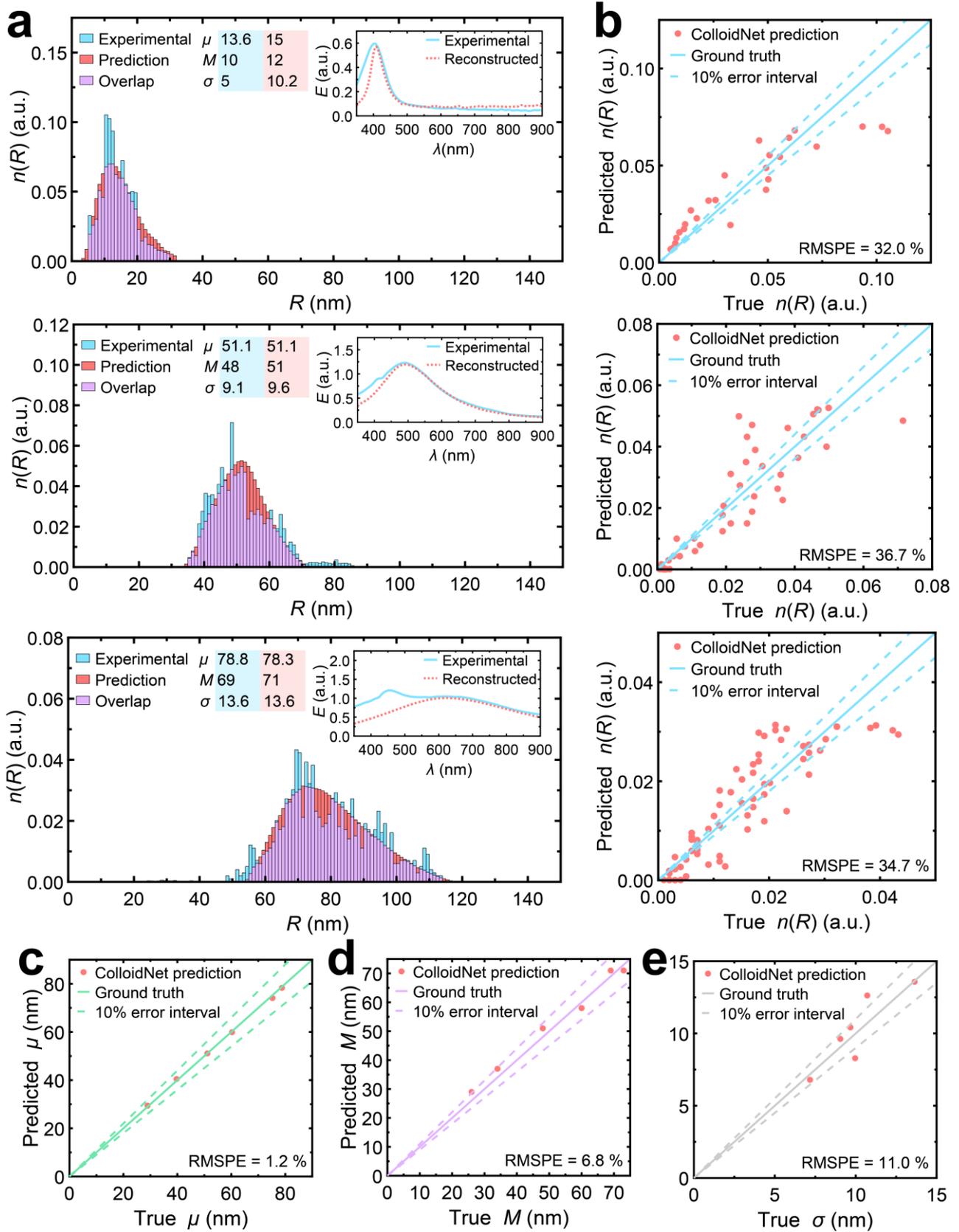

**Figure 5.** Comparison of NP size distribution characterization by SEM and "ColloidNet". (a) Experimental and predicted NP size distributions and their statistical distribution parameters for samples "F2", "E3", and "E6", the inset depicts experimental $E(\lambda)$ spectra and *dipole* spectra reconstructed from the predicted NP distributions and their volume filling fractions. (b) True-predicted value plots for distributions of (a) with the estimated RMSPE values and 10% absolute error intervals (dashed lines). True-predicted value plots for mean NP sizes ($\mu$), modes ($M$), and standard deviations ($\sigma$) and their respective RMSPE values for distributions of colloid series "E" are depicted in (c), (d), and (e), respectively, with solid lines corresponding to the true value while dashed lines correspond to a 10% absolute error interval.

**Predictions of NP concentration**

Since "ColloidNet" predicts only the volume filling fraction $F$, additional post-processing is needed to convert it into useful parameters such as Ag mass ($C_M$) and NP ($C_{NP}$) concentrations as explained in **Supplementary Section S2.5**. Predictions of "F" batch NP colloid Ag concentration acquired from "ColloidNet" predictions are depicted in **Figure 6**. For "ColloidNet", the LSPR $E(\lambda)$ peak magnitude is the key factor for predicting both $C_M$ and $C_{NP}$, due to it being directly influenced by $F$ (eq. S7, S9), according to[46]. However, the influence of the size distribution $n(R)$ becomes significant when computing $C_{NP}$, as it is directly impacted by NP size (eq. S9).

The literature described a relation between the intensity of $E(\lambda)$ at the interband wavelength which is 250 nm for Ag and the overall concentration of the material in the colloid[102,103]. Therefore, the dependence of $E(\lambda)$ magnitude at the interband together with the maximum position of LSPR on the mean NP radius is depicted in **Figure 6a**. The slightly increasing interband trend (**Figure 6a**) can be directly addressed to the stepwise Ag precursor addition at each NP synthesis. The interband $E(\lambda)$ is indicative of the overall Ag present in the solution ($C_M$), not just the Ag which is in NP form and follows the expected $AgNO_3$ content trend (**Figure 6b**) computed based on synthesis conditions tabulated data in **Table S1**. Moreover, the predicted $C_M$ is in the very same absolute mg/L level as expected based on the used chemical synthesis recipe and confirmed by the independent atomic mass spectroscopy measurement.

Finally, $C_{NP}$ can be computed from $C_M$ (**Figure 6c**) and compared with an expected value, which is computed from the synthesis conditions assuming the silver precursor is uniformly distributed among all NPs[51]. This is described in more detail in **Supplementary Section S2.5**. It is important to note that expected $C_{NP}$ values are computed assuming the NPs are monodisperse, while predicted $C_{NP}$ values have the distribution predicted by "ColloidNet". It is evident that dilutions of the colloids in each growth step play a key role in the decreasing trend of both predicted and expected $C_{NP}$ and are the ultimate cause of the decrease of $E(\lambda)$ relating to the LSPR peak as depicted in **Figure 6a** and **Figure S7a**. This is mainly because the decrease in the number of NPs cannot be compensated by the increase of the scattering component (**Figure S6a**), causing a net decrease in $E(\lambda)$. A very similar LSPR peak trend (**Figure S7b**) was observed by others as well [8]. While the decreasing $C_{NP}$ may seem like a quirk of the seeded growth synthesis method, dilution of the colloid used as a seed solution for the next growth step is paramount in order to ensure adequate growth of NPs – otherwise, the NPs are observed to hardly grow at all, even after multiple growth steps[51]. The NP size estimations in the seeded-growth synthesis often assume 100% consumption of the Ag precursor[51] while in our case around 40% provided a close result in the semilogarithmic plot depicted in **Figure 6c**. A closer to unit consumption might be expected if the NP distribution was taken into account while here size distribution was not estimated at all. But the expected and experimentally obtained tendencies hold and it for the last time here confirms that the DNN predicted Ag concentrations are trustworthy.

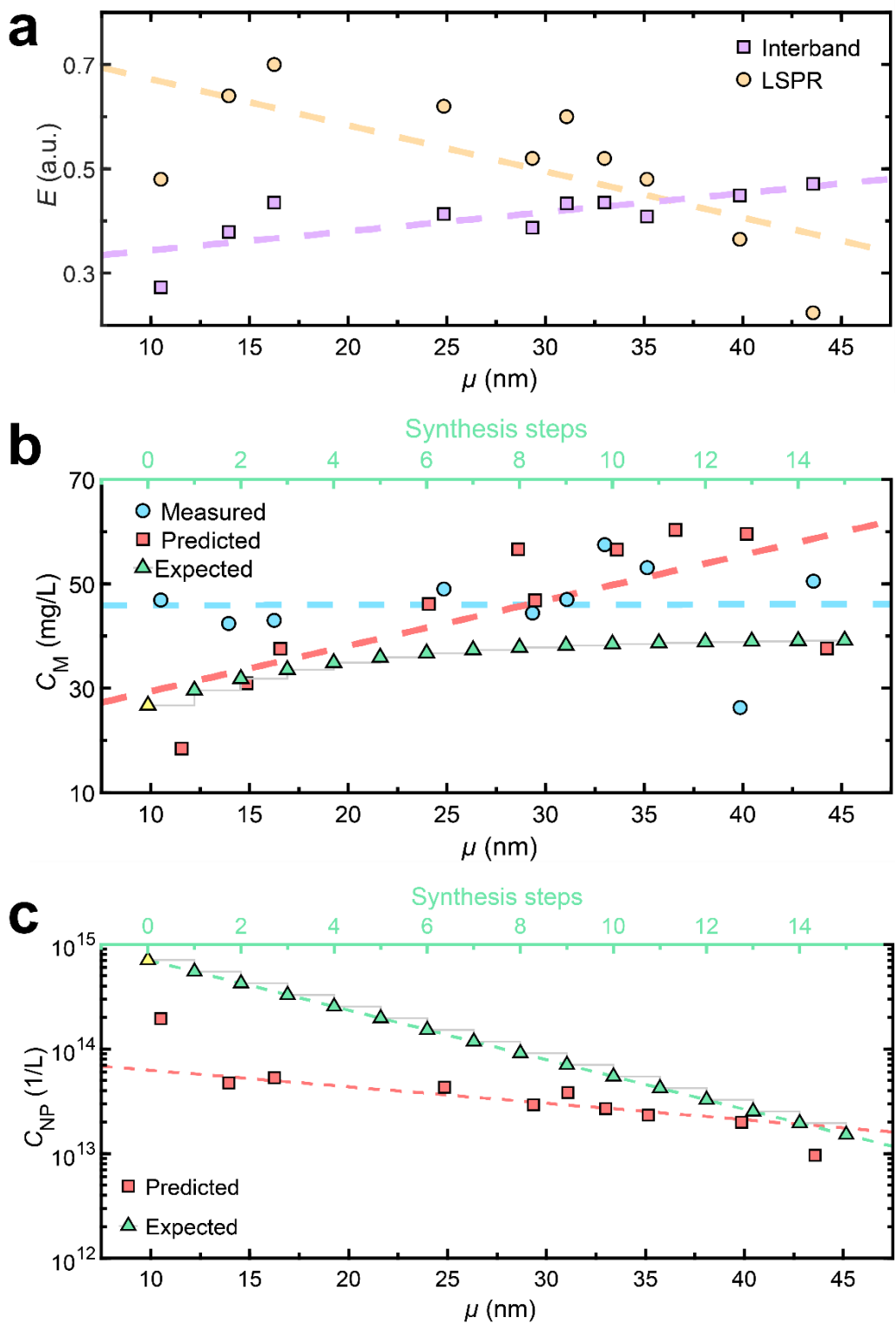

**Figure 6.** Concentration estimates for selected colloids of the "F" batch. (a) Trends of experimental $E(\lambda)$ of the interband transition (at 250 nm) and the LSPR peak found in **Figure S7a**. (b) Experimental *vs.* predicted *vs.* expected mass concentration ($C_M$). (c) "ColloidNet" predicted NP concentration ($C_{NP}$) *vs.* expected NP concentration assuming monodisperse NPs, 8 nm seeds, and 40% reaction efficiency with seed concentration $7.088·10^{14}$ and Ag density 10.49 g/cm³. The first "expected" values in (b) and (c) indicated in yellow are NP seeds. Lines are guides for the reader's eye.

**Limitations of the proposed method**

It is noteworthy that the performance of the proposed method can be affected by factors influencing the extinction spectra provided to the tandem DNN. In and of itself, "DipoleNet" is not affected by any factors affecting the input $E(\lambda)$ spectra – it simply recognizes and extracts the *dipole* component of the spectrum. However, those factors which affect the $E(\lambda)$ spectra also affect the *dipole* components extracted by "DipoleNet". On the other hand, predictions (both NP size and concentration) made by "ColloidNet" are influenced by factors affecting the input $E(\lambda)$ spectra. If the input spectrum is redshifted for any reason, then the NPs will be predicted to be larger than their actual size. Redshifting can occur due to NPs being placed in a host matrix (liquid) with a larger refractive index, having a thick coating of surfactants, or aggregation of NPs. Aggregation not only redshifts the spectrum but also decreases its magnitude. This will cause "ColloidNet" to evaluate the NP concentration as smaller than it actually is. All of these factors are described in more detail in **Supplementary Section S1.5.**

## Conclusions

In this work, we demonstrated the capability of a deep tandem neural network system to perform accurate silver nanoparticle size distribution and concentration predictions using UV-Vis extinction spectra as input.
Automated identification of the dipole peak contribution with the first DNN helped to accurately predict the NP size distribution with the second DNN in the wide range from 1 nm up to 150 nm retaining down to *ca.* 1.2% root mean square percentage error predicting mean nanoparticle size.
The tandem DNN predictions allowed for attaining accurate silver mass and nanoparticle concentration tendencies and their estimates throughout the seeded-growth synthesis steps were confirmed by the growing extinction at the interband and alternative atomic absorption measurement.
The extinction spectra computed according to predicted nanoparticle size distributions and volume-filling fractions matched the dipolar component of the experimental extinction spectra extremely well, further indicating an accurate prediction.
It was demonstrated that high accuracy and 11.3 ms per sample prediction rate can be achieved without any spectral data processing before providing it to the tandem DNN, making the proposed approach a good candidate for real-time nanoparticle colloid characterization under both laboratory and industrial conditions.
It was verified that the plasmonic NP size limit validity for the modified Maxwell-Garnett-Mie effective medium theory is much larger than previously thought and that the theory can be reliably used to estimate the size distribution of NPs based on their optical extinction dipole peak up to 150 nm in radius.

## Author Contributions

Conceptualization and methodology – T.K., T.T.; Investigation – T.K., N.K., A.T., L.V.; Data analysis – T.K., N.K., A.T., T.T.; Writing – original draft T.K., T.T., Writing – review & editing; T.K., N.K., A.T., S.T., T.T., Supervision – T.T.; Funding acquisition – T.T.

## Notes

The authors declare no competing financial interest.

## Supporting Information

The Supporting Information is available free of charge at: xx.xxxx/xxxxx.xxxxxxx.

- Details of Ag NP synthesis, volumes and concentrations of Ag NP precursors (Table S1), description of SEM micrograph analysis, example of SEM analysis (Figure S1), description of DLS measurements, description of XRD measurements, description of MMGM effective medium, parametrization of training and validation data, ranges of data parameters (Table S2), strategy of hyperparameter optimization, list of hyperparameters (Table S3), discussion about statistics of experimental data, statistical parameters of experimental data (Figure S2), discussion about DLS results, DLS results (Figure S3), discussion about XRD results, XRD results (Figure S4), XRD diffractogram parameters (Table S4), XRD crystallite size analysis (Table S5), discussion about the MMGM medium, comparison of MMGM and Mie theory results (Figure S5), comparison of scattering and absorbance (Figure S6), discussion about prediction of concentration, LSPR amplitude trends (Figure S7), discussion about mean size prediction trends for "C" NP batch (Figure S8), predictions of mean size for all investigated colloids (Figure S9, PDF)

# Acknowledgements

The authors acknowledge having received funding from the Research Council of Lithuania (LMTLT), agreement no. S-MIP-23-93.

Special thanks go to Dr. Karine Mougin from the Institut de Science des Matériaux de Mulhouse IS2M UMR for the technical assistance with the DLS measurements.

# Data Availability Statement

Experimental data is freely available at "Zenodo": https://doi.org/10.5281/zenodo.11091437. Other data can be made available after a reasonable request to the corresponding authors.

# Supplementary Information

# Deep Learning Methods for Colloidal Silver Nanoparticle Concentration and Size Distribution Determination from UV-Vis Extinction Spectra

Tomas Klinavičius[a*], Nadzeya Khinevich[a], Asta Tamulevičienė[a,b], Loic Vidal[c], Sigitas Tamulevičius[a,b], and Tomas Tamulevičius[a,b*]

[a]Institute of Materials Science of Kaunas University of Technology, K. Baršausko St. 59, LT-51423, Kaunas, Lithuania

[b]Department of Physics, Kaunas University of Technology, Studentų St. 50, LT-51368, Kaunas, Lithuania

[c]Institut de Science des Matériaux de Mulhouse IS2M UMR 7361, 15 rue Jean Starcky, F 68100 Mulhouse, France

*Corresponding authors Tel: +370 (37) 313432. T. Klinavičius: tomas.klinavicius@ktu.lt; T. Tamulevičius: tomas.tamulevicius@ktu.lt

# S1. Supplementary Methods

## S1.1 Detailed Description of Ag NP Synthesis

First, seeds were generated in the solution. The aqueous solution containing defined volumes of trisodium citrate (TC, with concentration $C_{TC}$) and tannic acid (TA, $C_{TA}$), was heated to 100°C and maintained at that temperature for 15 minutes under vigorous stirring. After that, a defined volume of silver nitrate (SN, $C_{SN}$) was injected. Then the NP seeds were left to form.

Second, NPs were synthesized by growing the seeds. NPs were grown by mixing a defined volume of seed solution with 16.5 ml of de-ionized water and heated to 90°C, then defined volumes of $C_{TC}$, $C_{TA}$, and $C_{SN}$ were sequentially injected, and synthesis lasted for 30 minutes. The produced Ag NP colloid was used as a seed colloid for the next synthesis and the growth process was repeated again (for each colloid the growth process was the same).

According to Bastús[1], the synthesized NPs are electrostatically stabilized by TC in water. Therefore, it can be expected that the NPs are covered in a thin coating of TC weakly bound to the NPs. The coating is expected to be sub-nanometer in thickness due to the small size of the TC molecule. However, due to the limited spatial resolution of the scanning electron microscope (SEM) micrographs (*ca.* 1.2 nm according to the manufacturer datasheet), the coating would not have a significant impact on the sharpness of the NP edges. Regardless, the coating was removed by ethanol before the deposition of NPs on a silicon substrate for SEM imaging.

Reagent volumes and concentrations used for the synthesis of both seeds and NPs are summarized in **Table S1** where seeds are denoted as the "0" samples, other samples are sequential according to their order of growth with letters "A"-"F" identifying different batches.

**Table S1.** Volumes and concentrations of materials used in seed growth (0) and further synthesis steps of different Ag NP batches ("A"-"F") along with corresponding statistical parameters of pre-produced NP size distributions. TA – tannic acid, TC – trisodium citrate, SN – silver nitrate along with statistical ($\mu$ – mean NP radius, $\sigma$ – standard deviation of NP radius, $R^2$ – goodness of fit) and ordering characteristics of empirical UV-Vis spectra (**Figure 2d**) as well as NP radius distributions (**Figure 2e**)

| Batch | No. | Synthesis components | | | | | Size distribution parameters | | | $R^2$ of fit | Sample order |
|---|---|---|---|---|---|---|---|---|---|---|---|
| | | Water (ml) | TA (ml, mM) | TC (ml, mM) | SN (ml, mM) | Previous Solution (ml) | $\mu$ (nm) | $\sigma$ (nm) | $\sigma$ (% of $\mu$) | | |
| A | A0 | 76 | 4, 2.5 | 20, 25 | | - | - | - | - | - | - |
| | A1 | 16.5 | 1.5, 2.5 | 0.5, 25 | 1, 25 | 19.5 | 12.8 | 3.1 | 24.2 | 0.90 | 2 |
| | A2 | | | | | | 15.7 | 5.3 | 33.7 | 0.96 | 5 |
| | A3 | | | | | | 24.1 | 5.9 | 24.4 | 0.94 | 10 |
| | A4 | | | | | | 27.5 | 5.8 | 21.1 | 0.93 | 14 |
| | A5 | | | | | | 30.9 | 3.2 | 10.4 | 0.93 | 19 |
| | A6 | | | | | | 35.7 | 4.9 | 13.7 | 0.93 | 25 |
| B | B0 | 60 | 20, 2.5 | 20, 25 | | - | - | - | - | - | - |
| | B1 | 16.5 | 1.5, 2.5 | 0.5, 25 | 1, 25 | 19.5 | 19.5 | 4.5 | 23.1 | 0.94 | 7 |
| | B2 | | | | | | 21.3 | 6.0 | 28.2 | 0.97 | 8 |
| | B3 | | | | | | 26.6 | 6.7 | 25.2 | 0.96 | 12 |
| | B4 | | | | | | 29.6 | 4.4 | 14.9 | 0.98 | 17 |
| | B5 | | | | | | 37.3 | 5.6 | 15.0 | 0.95 | 27 |
| | B6 | | | | | | 41.0 | 5.5 | 13.4 | 0.96 | 33 |
| C | C0 | 79 | 1, 100 | 20, 25 | 1, 25 | - | - | - | - | - | - |

| | | | | | | | | | | |
|---|---|---|---|---|---|---|---|---|---|---|
| | C1 | | | | | 19.5 | 26.7 | 3.5 | 13.1 | 0.96 | 13 |
| | C2 | | | | | | 31.1 | 6.7 | 21.5 | 0.97 | 21 |
| | C3 | 16.5 | 1.5, 2.5 | 0.5, 25 | | | 33.4 | 8.0 | 24.0 | 0.96 | 23 |
| | C4 | | | | | | 40.2 | 7.4 | 18.4 | 0.97 | 32 |
| | C5 | | | | | | 49.9 | 9.9 | 19.8 | 0.90 | 36 |
| | C6 | | | | | | 58.8 | 10.3 | 17.5 | 0.91 | 39 |
| D | D0 | 78.5 | 1.5, 100 | 20, 25 | | - | - | - | - | - | - |
| | D1 | | | | 1, 25 | 19.5 | 29.8 | 6.0 | 20.1 | 0.98 | 18 |
| | D2 | | | | | | 44.0 | 9.6 | 21.8 | 0.95 | 35 |
| | D3 | 16.5 | 1.5, 2.5 | 0.5, 25 | | | 54.3 | 11.3 | 20.8 | 0.92 | 38 |
| | D4 | | | | | | 67.2 | 12.1 | 18.0 | 0.95 | 41 |
| | D5 | | | | | | 86.3 | 13.5 | 15.6 | 0.93 | 44 |
| | D6 | | | | | | 107.6 | 18.1 | 16.8 | 0.87 | 45 |
| E | E0 | 77.5 | 2.5, 100 | 20, 25 | | - | - | - | - | - | - |
| | E1 | | | | 1, 25 | 19.5 | 28.9 | 7.2 | 24.9 | 0.97 | 15 |
| | E2 | | | | | | 39.7 | 10.0 | 25.2 | 0.91 | 29 |
| | E3 | 16.5 | 1.5, 2.5 | 0.5, 25 | | | 51.1 | 9.1 | 17.8 | 0.93 | 37 |
| | E4 | | | | | | 60.3 | 9.7 | 16.1 | 0.91 | 40 |
| | E5 | | | | | | 75.3 | 10.7 | 14.2 | 0.90 | 42 |
| | E6 | | | | | | 78.8 | 13.6 | 17.3 | 0.88 | 43 |
| F | F0 | 79.5 | 0.5, 2.5 | 20, 25 | 1, 25 | - | - | - | - | - | - |
| | F1 | | | | | | 10.5 | 3.1 | 29.5 | 0.95 | 1 |
| | F2 | | | | | | 13.6 | 5.0 | 36.8 | 0.94 | 3 |
| | F3 | | | | | | 13.9 | 4.8 | 34.5 | 0.94 | 4 |
| | F4 | | | | | | 16.2 | 5.0 | 30.9 | 0.98 | 6 |
| | F5 | | | | | | 21.7 | 5.7 | 26.3 | 0.97 | 9 |
| | F6 | | | | | | 24.8 | 4.4 | 17.7 | 0.95 | 11 |
| | F7 | | | | | | 29.3 | 2.6 | 8.97 | 0.93 | 16 |
| | F8 | 16.5 | 0.25, 2.5 | 0.1, 25 | 0.25, 25 | 58.5 | 28.1 | 3.5 | 12.5 | 0.98 | 20 |
| | F9 | | | | | | 33.0 | 3.6 | 10.9 | 0.98 | 22 |
| | F10 | | | | | | 35.1 | 2.9 | 8.3 | 0.99 | 24 |
| | F11 | | | | | | 35.8 | 2.8 | 7.8 | 0.99 | 26 |
| | F12 | | | | | | 39.3 | 2.4 | 6.1 | 0.99 | 28 |
| | F13 | | | | | | 39.9 | 3.8 | 9.5 | 0.98 | 31 |
| | F14 | | | | | | 39.7 | 2.7 | 6.8 | 0.99 | 30 |
| | F15 | | | | | | 43.6 | 2.8 | 6.4 | 0.99 | 34 |

## S1.2 Detailed Description of SEM Micrograph Analysis

An adaptive thresholding algorithm[2] was used to account for the inhomogeneity of image brightness. The scale of each micrograph was determined by automatically determining the length and reading the text (numeric value and units) of the scale bar – this allowed to avoid user error in determining the scale of the micrograph. Binarization of the image was performed according to this local threshold value. In order to eliminate possible binarization artifacts due to uneven brightness of spherical-like NPs resulting in donut-shaped image regions with holes they were filled in[3]. The watershed segmentation employing the Fernand Meyer algorithm[4] was used to separate NP regions that were stuck together after binarization in order to estimate the size of a single NP instead of several or an entire aggregate, this increased the reliability of the radius data for statistics. In order to determine the effective radius of NPs, their post-segmentation area in the image was equated to the area of a circle with the effective radius. Radii smaller than 1% of the scale bar were discarded from the statistical analysis in each micrograph because their estimates were found to be unreliable due to the SEM resolution limit. Each detected NP is numbered according to its detection order, this allows to potential outlier identification. NPs touching the boundary of the micrographs were removed from the analysis results in order to have only particles of full size. Probability density functions of effective radius distributions and statistical parameters of said distributions were computed from the effective radius data. The key steps of the used image analysis algorithm are depicted in **Figure S1** by using a fragment of a single SEM micrograph.

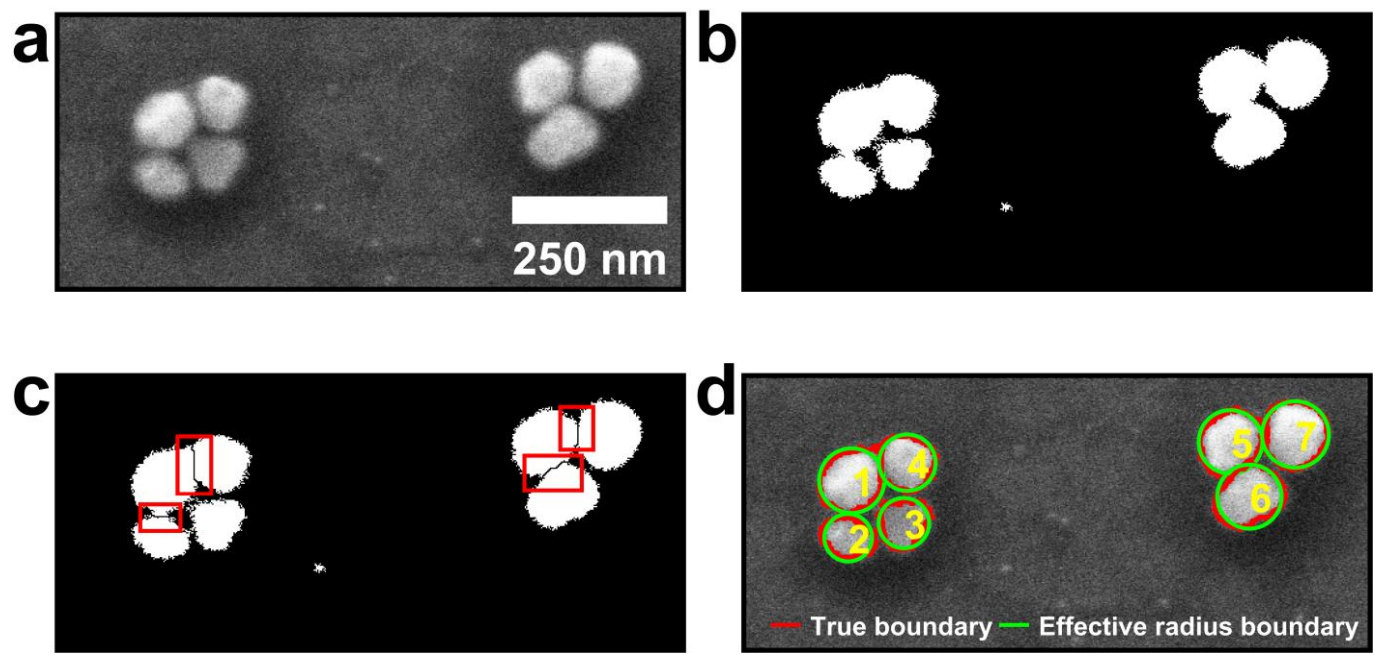

**Figure S1.** Key steps of the SEM micrograph analysis algorithm. (a) Determination of scale of the input image, performed directly from the scale bar, the scale bar here was added only for reference. (b) Binarization of the image based on adaptive thresholding, in order to separate the NPs from the background. (c) Segmentation of NP clusters into individual NPs using watershed transformation. (d) The fully processed output image with true and effective NP boundaries along with the NP number.

## S1.3 Detailed Description of NP Size Measurement by Dynamic Light Scattering

The hydrodynamic particle radius and its distribution were measured using dynamic light scattering (DLS) technique. Measurements were carried out on a "Zeta Sizer Ultra" particle size analyzer (Malvern Panalytical). The scattering angle was fixed at 90°. 2 ml of colloidal solution was used for the measurement in a disposable cuvette. Before the measurement, the colloidal solution was agitated using an ultrasound bath to avoid aggregation of the NPs in the measuring solution.

## S1.4 Detailed Description of NP Crystallinity Measurements by X-Ray Diffractometry

X-ray diffractometer "D8 Discover" (Bruker AXS GmbH) was used to perform X-ray diffraction (XRD) measurements and estimate the crystallinity of Ag NPs. Cu K$\alpha_1$ ($\lambda$ = 1.5406 Å) radiation was used as the X-ray source, utilizing a parallel beam-focusing geometry. XRD spectra were measured in the 30°-90° ($\theta$-$2\theta$) with a step size of 0.012° using a silicon strip LynxEye detector. Data was analyzed using the DIFFRAC.EVA software (Bruker). The crystallite size ($C$) was computed using the Scherrer equation[5].

## S1.5 Detailed Description of Modified Maxwell-Garnett-Mie Effective Medium Theory

The permittivity of materials composed of weakly interacting sub-wavelength particles disorderly dispersed in a host medium can be described by the Maxwell-Garnett effective medium theory (EMT)[6]. If the particles are spherical or almost spherical, such materials can be more accurately described by Maxwell-Garnett-Mie EMT[7], because this theory takes into account the polarizability of the particle derived by Mie theory. However, this model also assumes monodisperse inclusions. In general, such materials are composed of particles of various sizes. The modified Maxwell-Garnett-Mie (MMGM)[8] EMT is able to account for the probability distribution of particles dispersed within a defined size range. For noble metal NPs the permittivity can be computed from:

$$\frac{\varepsilon_{eff} - \varepsilon_m}{\varepsilon_{eff} + 2\varepsilon_m} = \frac{3i\lambda^3}{16\pi^3 \varepsilon_m^{3/2}} \frac{F}{\mu} \int_{R_{min}}^{R_{max}} P(R)\, a_1(\varepsilon_{NP}, R)\, dR \quad \text{(eq. S1)}$$

Here $\varepsilon_{eff}$ is the effective permittivity, $\varepsilon_m$ is the permittivity of the host medium, $\lambda$ is the vacuum wavelength of light, $F$ is the volume filling fraction of the particles in the host medium, $\mu$ is the mean size of the particles that are distributed according to the probability density law $P(R)$, $a_1(\varepsilon_{np}, R)$ is the first electric Mie coefficient (which accounts for NP permittivity and size) while $R_{min}$ and $R_{max}$ are the lower and upper bounds of the NP distribution range.

As expressed by (eq. S1), the permittivity of the effective medium is directly affected by that of the host medium of the NPs. A host medium with a larger permittivity (or refractive index) will redshift the LSPR peak[3]. The NP having a coating of significant thickness can have a similar effect[9]. The NPs produced in this work were electrostatically stabilized by TC in water[1]. In[9], no shift in the SPR peak of nanoparticles in water functionalized by TC was observed when comparing them to non-stabilized particles. This is likely due to the relatively low concentration of TC in the solution, where the molality of TC in the growth solution was on the order of $10^{-3}$ mol/kg, and the change in refractive index was observed only when molality was around 0.2 mol/kg[10]. TA had similarly miniscule levels of concentration in the colloid as TC. Finally, the MMGM EMT assumes NPs are sufficiently dilute and well distributed in the host medium, hence no aggregation of NPs, which may occur in unsonicated real samples, would be possible to describe. The aggregation of nanoparticles can significantly decrease the intensity of LSPR in absorbance spectra. Additionally, a minor shift toward longer wavelengths and broadening of the LSPR peak may also be observed[11].

Y. Battie *et al.* in[8] claim that MMGM theory can only be used for noble metal NPs with a clear-cut condition of their radius being <25 nm due to the concept of an effective permittivity having a physical meaning only if scattering does not supersede absorption. However, the formula (eq. S1) does not have any built-in limitation for using it outside the designated size range. Moreover, D. Werdehausen *et al.* in[12] have shown that the transition from EMT being valid to being invalid is in fact gradual. They have shown the partial validity of EMT for NP radius values at least double that of what is claimed in[8], making the limit where it is no longer applicable somewhat blurred. Therefore, it cannot be unambiguously stated that EMT is invalid for NPs larger than 25 nm.

Conversely, it should be noted that NPs have a certain lower size limit. The gap between lone atoms and NPs is bridged by what are known as atom clusters[13]. For metal NPs, their permittivity generally depends on their size, especially for small NPs, while the electronic band structure remains continuous. On the other hand, clusters have a key difference – their band structure is discrete, like that of a molecule, and in essence quantum in nature. Classical approaches like Mie theory are considered to work with non-quantum objects. Intrinsic (non-quantum) effects were taken into account when considering permittivity according to[8]. The transition from quantum to classical is usually considered to happen at *ca.* 1 nm radius[13–16]. Some other sources claim smaller sizes, such as about 40 atoms specifically for silver[17]. In this work, the limit of clusters was considered to be 1 nm in radius, hence it is the smallest radius value chosen to be predicted by the tandem network. It is also in line with the resolution of the field emission SEM used in the research.

## S1.6 Log-normal Distributions and Extinction Spectra Parameters for Generation of Training and Validation Data

Both *full* and *dipolar* extinction $E(\lambda)$ spectra ranged from 350 nm to 900 nm with an increment of 2 nm. Intermediate boundary values of $FWHM_L$ were determined by linearly interpolating between the boundary values of FWHM$_L$ for the minimum and maximum values of $M_L$ using the equations (eq. S2) and (eq. S3) respectively. These equations were derived using the equation for a line going through 2 points.

$$\text{FWHM}_{L\,\text{lower}} = 0.0762 M_L + 1.6190 \quad \text{(eq. S2)}$$
$$\text{FWHM}_{L\,\text{upper}} = 0.1905 M_L + 19.0476 \quad \text{(eq. S3)}$$

$M_L$ and FWHM$_L$ are not commonly used to define the log-normal distribution – even an ontology dedicated to distributions offers no parametrization based on $M_L$ and FWHM$_L$[18]. Therefore, a relation between ($M_L$, FWHM$_L$) and ($\mu_L$, $\sigma_L$) was derived using the symbolic toolbox of "MATLAB". The transformation is defined by equations (eq. S4) and (eq. S5):

$$\mu_L = \ln(M_L) + \frac{\left(\ln(M_L) - \ln\left(FWHM_L + \sqrt{4M_L^2 + FWHM_L^2}\right) + \ln(2)\right)^2}{2\ln(2)} \quad \text{(eq. S4)}$$

$$\sigma_L = -\frac{\ln(M_L) - \ln\left(FWHM_L + \sqrt{4M_L^2 + FWHM_L^2}\right) + \ln(2)}{\sqrt{2\ln(2)}} \quad \text{(eq. S5)}$$

These canonical log-normal distribution parameters $\mu_L$ and $\sigma_L$ were then used to originate the log-normal distributions using "MATLAB". As NPs grow their size distribution tends to become more dispersed[1]. To reflect this, for the largest value of $M_L$ the FWHM$_L$ range was selected to be larger. Ranges and iteration steps of both ($M_L$, FWHM$_L$) and ($\mu_L$, $\sigma_L$) are provided in **Table S2**.

**Table S2.** Parameters for generating training and validation data. $R$ – radius over which the log-normal distributions are defined, $M_L$ – mean of the log-normal distribution, FWHM$_L$ – full width at half maximum of the log-normal distribution

| Dataset | Parameter | Lower bound | Upper bound | Step |
|---|---|---|---|---|
| Log-normal distribution range | $R$ | 1 nm | 150 nm | 1 nm |
| Training | $M_L$ | 5 nm | 110 nm | 1 nm |
| | FWHM$_L$ ($M_L$ = 5 nm) | 2 nm | 20 nm | 1 nm |
| | FWHM$_L$ ($M_L$ = 110 nm) | 10 nm | 40 nm | 1 nm |
| | FWHM$_L$ (5 nm < $M_L$ < 110 nm) | Equation (S2) | Equation (S3) | 1 nm |
| | The amplitude of MMGM extinction | 0.1 | 2.5 | 0.01 |
| Validation | $M_L$ | 5 nm | 110 nm | Random |
| | FWHM$_L$ ($M_L$ = 5 nm) | 2 nm | 20 nm | Random |
| | FWHM$_L$ ($M_L$ = 110 nm) | 10 nm | 40 nm | Random |
| | FWHM$_L$ (5 nm < $M_L$ < 110 nm) | Equation (S2) | Equation (S3) | Random |
| | Amplitude of MMGM extinction | 0.1 | 2.5 | Random |

## S1.7 Use of Training and Validation Data to Set Optimal Hyperparameters

For each DNN, training data and validation data consisted of inputs and outputs correspondingly coupled into data pairs. Training was conducted using a minibatch gradient descent process for both DNNs, therefore a number of data pairs were used in each step of both training and validation for both DNNs. During hyperparameter optimization (HPO), and subsequently – the final training using optimal hyperparameters, input members of data pairs were given to the DNN and the DNN provided an output based on those inputs. Then, the outputs of the DNN were compared to the output members of the data pairs, and a root-mean-squared error was computed for adjusting the weights and biases of the DNNs using the root mean square propagation algorithm for backpropagation.

Training data was used to train the DNNs, while validation data was used to monitor training performance during HPO. Testing data was used to evaluate the performance of the final, post-HPO, model, trained using optimal hyperparameter values and a complete training dataset. HPO parameters along with user-set parameters are provided in **Table S3**.

For "DipoleNet", training and validation data pairs consisted of numerically generated *full* $E(\lambda)$ spectra for the inputs and numerically generated *dipolar* $E(\lambda)$ spectra for the outputs. Mie theory was used to generate *full* spectra and TMM using MMGM medium was used to generate *dipole* spectra. Testing data inputs consisted of measured $E(\lambda)$ spectra for the inputs and best-fit dipole spectra, computed using TMM with MMGM medium, for the outputs.

Similarly, training and validation data for "ColloidNet" were composed of numerically generated *dipolar* spectra and corresponding numerical NP radius distributions with their *F* values. TMM using MMGM medium was used to originate the *dipole* spectra, and the size distributions with their *F* values were numerically generated and used to compute the permittivity of the MMGM medium[8]. For inputs of testing data, outputs of "DipoleNet" were used once it was confirmed that "DipoleNet" had achieved optimal performance. Meanwhile, outputs of testing data were experimentally derived NP distributions which correspond to measured spectra used for testing inputs of "DipoleNet" along with their measured concentrations. Separately using both DNNs one after the other completes the functionality of the "tandem" DNN, allowing the DNN system to take measured spectra as inputs and output predictions of what the NP distribution and its NP concentration should be.

**Table S3.** Optimizable and user-set hyperparameters for both DNNs and their optimal values. Range bounds of user-set parameters are denoted as "-".

| Network | Hyperparameter | Lower bound | Upper bound | Optimal value |
|---|---|---|---|---|
| DipoleNet | Size of the input layer | - | - | 276 |
| | Number of neurons in each hidden layer | 276 | 1500 | 1486 |
| | Size of the output layer | - | - | 151 |
| | Number of hidden layer blocks | 1 | 5 | 3 |
| | Leaky ReLU scale | - | - | 0.01 |
| | Size of mini-batch | 1 | 100 | 1 |
| | Initial learning rate | $1 \cdot 10^{-5}$ | $1 \cdot 10^{-2}$ | $2.5218 \cdot 10^{-5}$ |
| | Learning rate multiplier | 0.1 | 0.9 | 0.7235 |
| | Learning rate drop period | 3 | 15 | 11 |
| | Number of training epochs for HPO | - | - | 100 |
| | Validation frequency in iterations for HPO | - | - | 50 |
| ColloidNet | Size of the input layer | - | - | 276 |
| | Number of neurons in each hidden layer | 276 | 1500 | 995 |
| | Size of the output layer | - | - | 151 |
| | Number of hidden layer blocks | 1 | 5 | 2 |
| | Leaky ReLU scale | - | - | 0.01 |
| | Size of mini-batch | 1 | 50 | 12 |
| | Normalization constant (epsilon) | - | - | $10^{-5}$ |
| | Dropout probability | - | - | 0.5 |
| | Magnitude of L2 regularization | - | - | $1 \cdot 10^{-4}$ |
| | Initial learning rate | $1 \cdot 10^{-5}$ | $1 \cdot 10^{-2}$ | $6.0203 \cdot 10^{-4}$ |
| | Learning rate multiplier | 0.1 | 0.9 | 0.5276 |
| | Learning rate drop period | 3 | 15 | 6 |
| | Number of training epochs for HPO | - | - | 100 |
| | Validation frequency in iterations for HPO | - | - | 50 |

# S2. Supplementary Results & Discussion

## S2.1 Statistical Parameters of Experimental NP Size Distributions

The experimental NP size distribution parameters extracted from the scanning electron microscopy analysis are depicted in **Figure S2a**. The standard deviation ($\sigma$) of NP size tends to increase when their mean radius ($\mu$) increases. The exceptions are the most monodisperse batches "A" (red), "B" (Orange), and "F" (blue), where this trend is not observed. In **Figure S2b**, when $\sigma$ is expressed in units relative to $\mu$, this trend truly becomes noticeable – for each batch with progressively larger $\mu$, the decrease of $\sigma$ becomes less pronounced, while for largely monodisperse NPs it continues to decrease. Based on this, the monodisperse batches were grouped together and fitted with a least-squares purple line, while the polydisperse batches were grouped together and fitted with a gray line to better illustrate the trend.

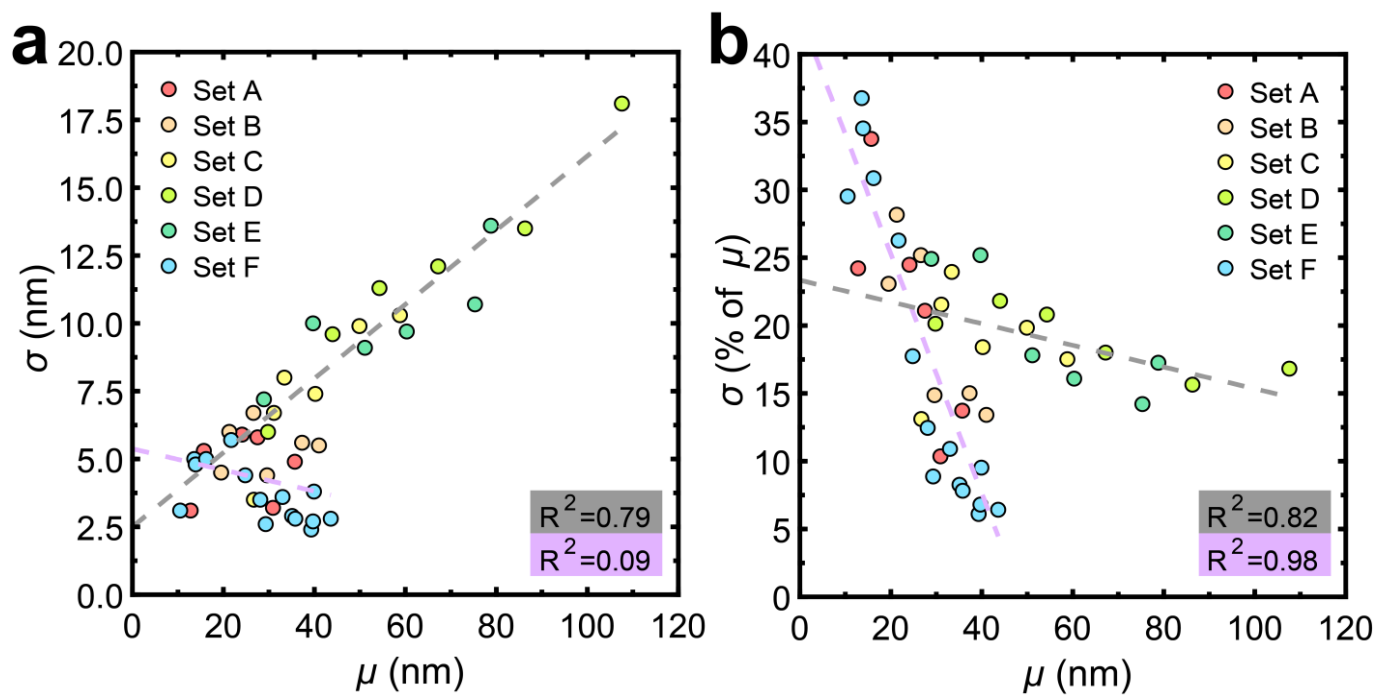

**Figure S2.** Mean ($\mu$) and standard deviation ($\sigma$) of experimental NP size distributions. (a) Dependence of standard deviation on mean size for different batches of NPs. (b) Dependence of standard deviation, expressed as a percentage of the mean, on mean size for different batches of NPs. Corresponding data is available in **Table S1** Truncated lines represent linear regression but in (a) it resembles guide for the reader's eye as $R^2$ values are low for sets "A" and "F" in (a).

## S2.2 Measurement Results of Ag NP Size Distribution by Dynamic Light Scattering

It was observed that for investigated samples DLS tends to significantly underestimate the mode of the NP size distribution while exaggerating the FWHM of the distribution, making it seem like the NPs are more polydisperse than they are according to SEM micrograph analysis, as evident in **Figure S3**. That is because the output radii of DLS are logarithmically spaced due to numerical methods used by the analysis software[19]. Similar issues when analyzing gold NPs were described by others[20], therefore NP size characterization by SEM was chosen for our work.

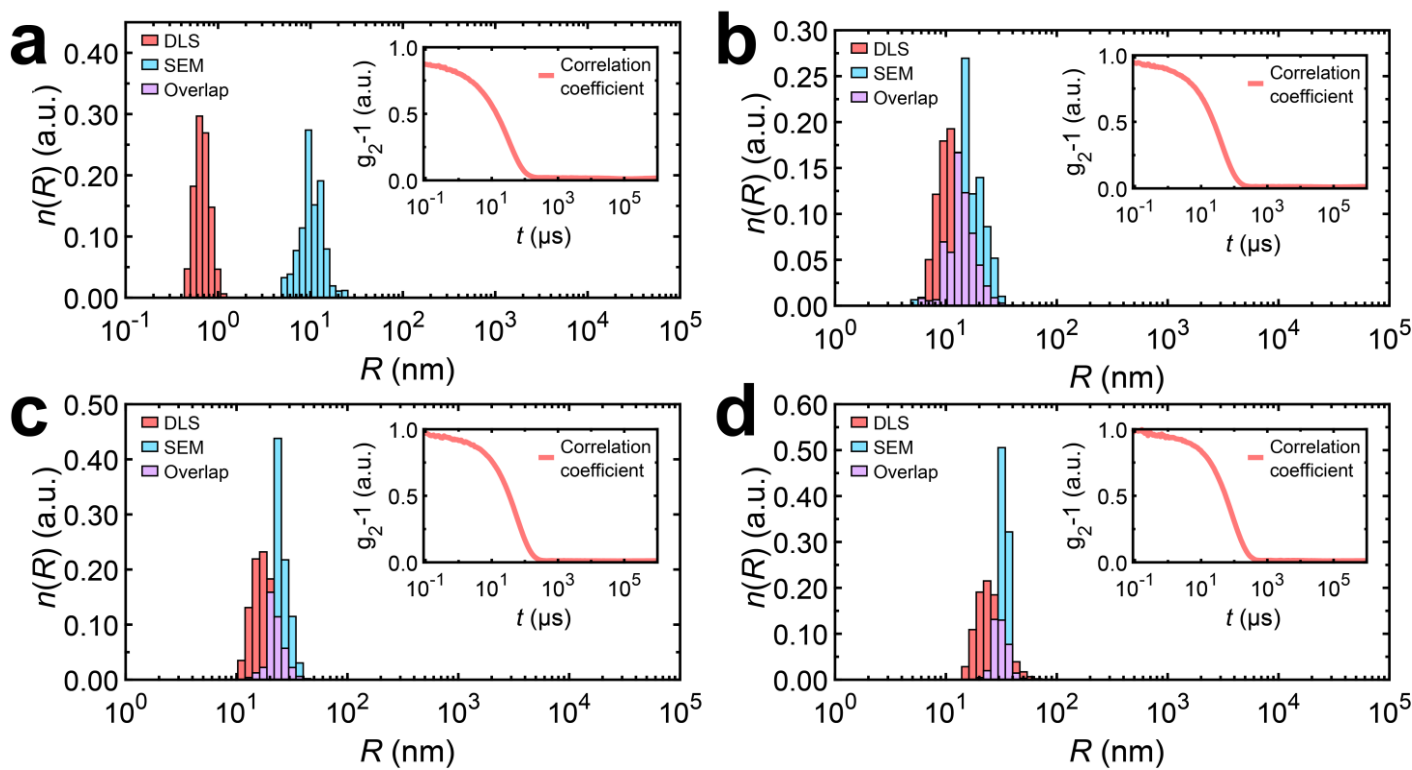

**Figure S3.** Comparison of Ag NP size distributions obtained by DLS and SEM with corresponding DLS correlogram. (a) Sample "F1". (b) Sample "F4". (c) Sample "F6". (d) Sample "F9".

## S2.3 Results of Crystallinity Measurements of Ag NPs by X-Ray Diffractometry

XRD diffractograms were measured for selected samples of Ag NPs, they are displayed in **Figure S4a**. The diffraction peaks were identified to be the crystal planes (111), (200), (220), (311), and (222) respectively, indicating a face-centered cubic lattice, which is characteristic of pure Ag. The peaks can be seen to correspond fairly well (**Table S4**) to the standard powder diffraction card for Ag of the Joint Committee on Powder Diffraction Standards (JCPDS 04-0783). Estimates of crystal lattice parameters were performed according to[21]. The weak intensity of the peaks shows poor crystallinity of the synthesized Ag NPs, indicating they are likely polycrystalline[22].

Crystallite size with its standard deviation, mean radius of the investigated sample with its standard deviation and also the ratio of crystallite size to the mean radius with the ratio's standard deviation are presented in **Table S5**. The standard deviation of the ratio is computed in accordance with the rules of error propagation[23]. Overall, **Figure S4b** shows a clear trend of crystallite size increasing along with the NP radius. Conversely, **Figure S4c** shows a reverse trend, proving that NPs end up becoming more polycrystalline as they grow. This is due to the synthesis method used, as other authors[22] obtained a similar trend.

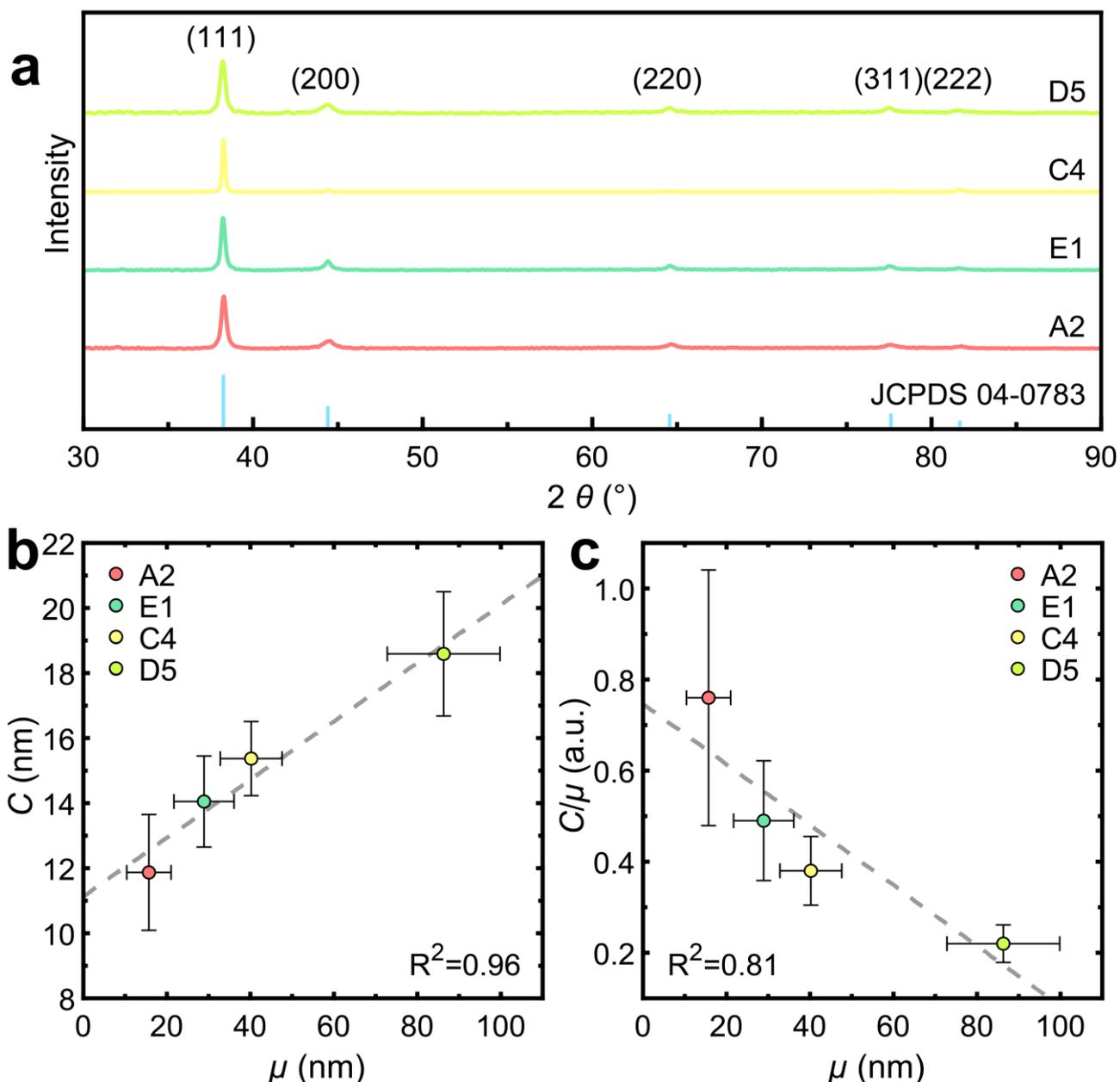

**Figure S4.** XRD analysis. (a) Several selected Ag NP samples, representing different effective radius values, XRD diffractograms with assigned diffraction peaks and the standard powder diffraction card for Ag (JCPDS 04-0783) (offset for clarity). (b) Dependence of crystallite size ($C$) on mean NP radius ($\mu$). (c) Dependence of ratio between crystallite size and mean NP radius on mean NP radius. All error bars are 1 standard deviation.

**Table S4.** XRD results. Miller indices ($hkl$), peak locations ($2\theta$), full width at half maximum of the peaks ($\beta$), estimates of interplanar spacings ($d_\theta$), estimates of lattice parameter ($a_\theta$), mean lattice parameter and standard deviation of lattice parameter estimates ($\Delta a$) for select samples

| Sample name | ($hkl$) | $2\theta$ (°) | $\beta$ (°) | $d_\theta$ (Å) | $a_\theta$ (Å) | $a$ (Å) | $\Delta a$ (Å) |
|---|---|---|---|---|---|---|---|
| A2 | (111) | 38.27 | 0.38 | 2.3497 | 4.0698 | 4.0734 | 0.0052 |
|    | (200) | 44.52 | 0.48 | 2.0331 | 4.0661 |  |  |
|    | (220) | 64.62 | 0.49 | 1.4411 | 4.0762 |  |  |
|    | (311) | 77.58 | 0.39 | 1.2295 | 4.0778 |  |  |
|    | (222) | 81.76 | 0.49 | 1.1769 | 4.0771 |  |  |
| E1 | (111) | 38.21 | 0.32 | 2.3531 | 4.0756 | 4.0789 | 0.0029 |
|    | (200) | 44.41 | 0.40 | 2.0383 | 4.0766 |  |  |

S10

|  |  |  |  |  |  |  |  |
|---|---|---|---|---|---|---|---|
|  | (220) | 64.58 | 0.39 | 1.4420 | 4.0786 |  |  |
|  | (311) | 77.50 | 0.38 | 1.2307 | 4.0819 |  |  |
|  | (222) | 81.64 | 0.38 | 1.1783 | 4.0818 |  |  |
| C4 | (111) | 38.22 | 0.31 | 2.3531 | 4.0756 | 4.0786 | 0.0023 |
|  | (200) | 44.38 | 0.33 | 2.0396 | 4.0791 |  |  |
|  | (220) | 64.58 | 0.30 | 1.4419 | 4.0782 |  |  |
|  | (311) | 77.49 | 0.37 | 1.2307 | 4.0819 |  |  |
|  | (222) | 81.74 | 0.39 | 1.1772 | 4.0779 |  |  |
| D5 | (111) | 38.26 | 0.24 | 2.3505 | 4.0712 | 4.0775 | 0.0048 |
|  | (200) | 44.44 | 0.30 | 2.0371 | 4.0742 |  |  |
|  | (220) | 64.58 | 0.25 | 1.4420 | 4.0786 |  |  |
|  | (311) | 81.62 | 0.33 | 1.2304 | 4.0808 |  |  |
|  | (222) | 77.52 | 0.29 | 1.1786 | 4.0828 |  |  |
| JCPDS 04-0783 | (111) | 38.12 | - | 2.3590 | 4.0859 | 4.0860 | 0.0020 |
|  | (200) | 44.28 | - | 2.0440 | 4.0880 |  |  |
|  | (220) | 64.43 | - | 1.4450 | 4.0871 |  |  |
|  | (311) | 77.47 | - | 1.2310 | 4.0828 |  |  |
|  | (222) | 81.54 | - | 1.1796 | 4.0863 |  |  |

**Table S5.** Mean NP radius ($\mu$), standard deviation of radius ($\Delta\mu$), mean crystallite radius ($C$), standard deviation of crystallite radius ($\Delta C$), ratio of mean crystallite radius and mean NP radius ($C/\mu$), and the standard deviation of the ratio ($\Delta(C/\mu)$) for select samples

| Sample name | $\mu$ (nm) | $\Delta\mu$ (nm) | $C$ (nm) | $\Delta C$ (nm) | $C/\mu$ | $\Delta(C/\mu)$ |
|---|---|---|---|---|---|---|
| A2 | 15.7 | 5.3 | 11.87 | 1.78 | 0.76 | 0.37 |
| E1 | 28.9 | 7.2 | 14.05 | 1.40 | 0.49 | 0.27 |
| C4 | 40.2 | 7.4 | 15.37 | 1.14 | 0.38 | 0.20 |
| D5 | 86.3 | 13.5 | 18.59 | 1.91 | 0.22 | 0.19 |

## S2.4 Validity of the Effective Medium Model

It can be observed that as the mean size of NPs increases, the $E(\lambda)$ spectrum gains more peaks while the previously existing peak redshifts (**Figure 2 d**). This occurs due to multiple modes of oscillation (called multipoles) being allowed to exist simultaneously in large particles due to retardation effects, which are explained by Mie theory[24], while only a single mode exists in small particles compared to the wavelength in the host medium (**Figure S5 a**, **Figure S6 b** red curve). What is more, these independent modes exist for both absorption and scattering, which are the constituent parts of $E(\lambda)$ (**Figure S5 a**). For sufficiently small NPs absorbance is larger than scattering by orders of magnitude, but the magnitude of scattering tends to increase more rapidly with size than absorbance (**Figure S6 a**) and eventually overtakes it.

For small NP sizes where absorption is the dominant ingredient of $E(\lambda)$ (*i.e.* **Figure S6 b** red curve), the quasi-static approximation[3] holds true and a dilute collection of such NPs can be homogenized into an effective medium[8]. The Maxwell-Garnett-Mie (MMGM) effective medium was used in this work. The MMGM medium can only account for the dipolar component of overall absorbance. However, when Ag NP radius reaches ~25 nm absorbance and scattering become equal in amplitude (indicated by an arrow in **Figure S6 a**). From that point forward, scattering begins to dominate in the $E(\lambda)$ spectrum (**Figure S6b** green and blue curves) and the quasi-static approximation is no longer valid. Therefore, the host medium with embedded Ag NPs inclusions can no longer be considered homogeneous for this size range, and said MMGM medium should no longer accurately describe the spectral response of the colloid. However, this is debatable, since the transition from effective medium validity to invalidity is not abrupt. According to[12], for NPs with a radius of *ca.* 50 nm effective medium theories like MMGM still predict their optical properties with a good deal of accuracy and only NPs larger than 100 nm in radius cannot be homogenized at all. We present a surprising case of the MMGM effective medium being seemingly valid for Ag NPs almost up to 150 nm in radius for predicting the shape of $E(\lambda)$ spectra (**Figure S5 b**). It might seem impossible, but the NP size distribution used to model the colloid clearly matches the experimentally derived NP size distribution (**Figure S5c**), indicating this unlikely claim is empirically true. Moreover, the modelling result obtained using MMGM effective medium also corresponds well with $E(\lambda)$ predictions made using Mie theory (**Figure S5 a**). For NPs with a radius <25 nm, where the MMGM theory holds[8], the filling fraction $F$ corresponds to an actual volume filling fraction (concentration) of NPs in the host medium and can accurately determine their absorbance spectrum. What is more, it directly corresponds to the amplitude of the overall $E(\lambda)$ signal. At that size range $E(\lambda)$ is composed almost exclusively of absorbance, which is what TMM using an MMGM medium actually computes. The amplitudes of $E(\lambda)$ and absorbance match well due to scattering being negligible. The amplitude of the absorbance dipole drops significantly for larger NPs, this is accurately predicted by the MMGM medium. Therefore, in order for the model spectrum amplitude to keep up with the experimental amplitude, $F$ has to be normalized taking into account the NP size-dependent absorbance contributions, essentially making it a free parameter of the MMGM model. Because of this, $F$ no longer directly describes NP concentration, but the part of the spectrum corresponding to the dipole resonance can still be computed and its amplitude can match that of the experimental spectrum, despite the physics not being represented with complete accuracy. Since DNNs need large quantities of training data with the magnitude of said data comparable to testing ("real world") data and DNNs themselves are not concerned about accurate physics internally, we find this to be a valid technique, especially since the shape of the spectrum is still related to the NP size distribution (**Figure S5c**). The conversion of $F$ into actual NP concentration is described in **Section S2.6.**

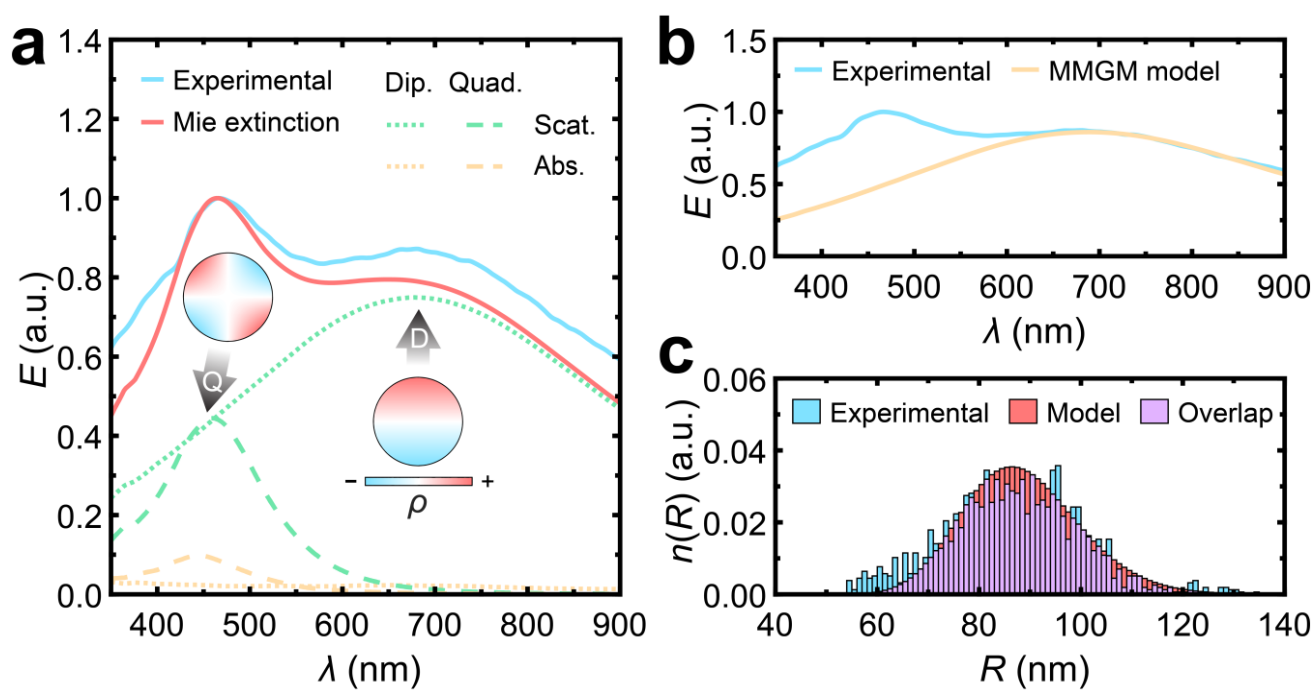

**Figure S5.** Comparison of Mie theory and MMGM effective medium theory. (a) Decomposition of Ag NP colloid "D5" ($\mu$ = 86.3 nm) $E(\lambda)$ spectrum, normalized to one, in terms of multipoles (dipole "D" and quadrupole "Q") of absorbance and scattering, showing the surface charge distribution ($\rho$) corresponding to each LSPR oscillation mode. (b) Fitting of the "D5" colloid normalized $E(\lambda)$ dipolar peak amplitude with TMM using the MMGM effective medium permittivity. (c) Experimental NP radius distribution of "D5" compared to the model distribution used for both Mie decomposition and the MMGM effective permittivity computation.

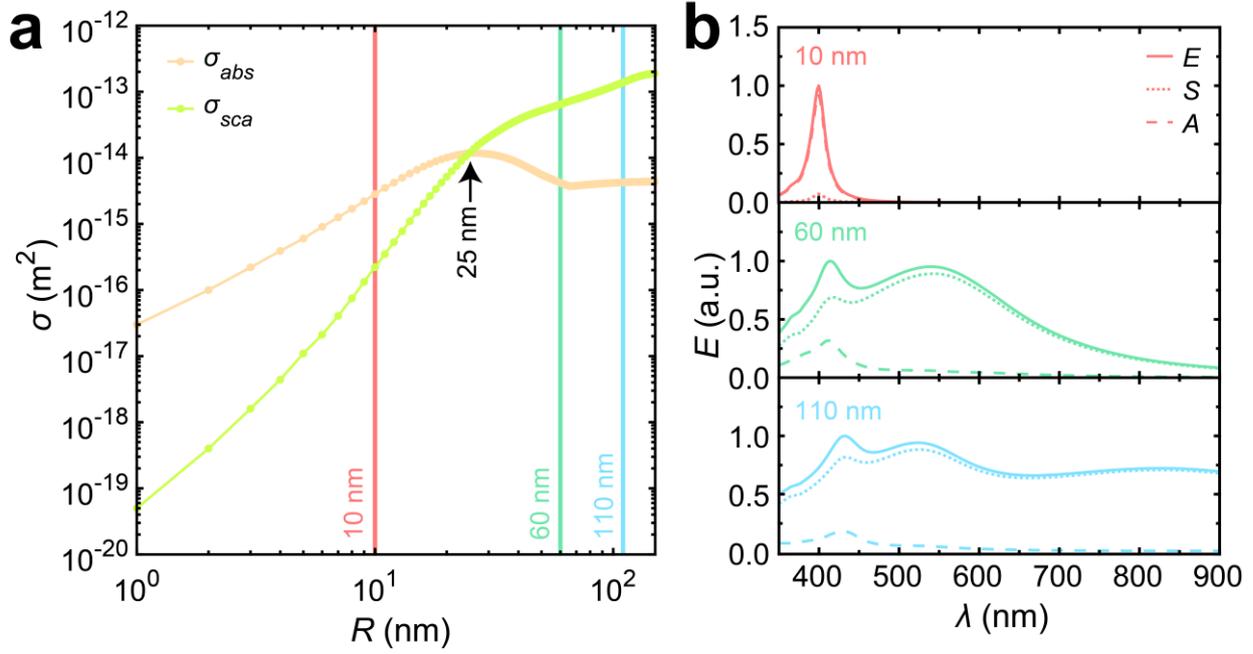

**Figure S6.** Comparison of absorbance ($\sigma_{abs}$) and scattering ($\sigma_{sca}$) cross-sections. (a) Logarithmic plot of maximum absorbance and scattering cross-sections for the dipole resonance (selected from the spectral range 350 nm – 900 nm) denoting the radius of equal amplitude indicated by the arrow. (b) $E(\lambda)$, normalized to one, of NPs with distinct radius values, denoted in (a), showing its constituent parts of multipolar scattering (point lines) and absorbance (dashed lines).

## S2.5 Ag Mass and NPs Concentration Estimation Using Predicted Filling Fraction

The filling fraction ($F$), despite being intuitive, is not a parameter that is directly comparable to measurement results like those of mass concentration (mg/L). The concentration of NPs (1/L) is another parameter that may be of use. Therefore, a relation between $F$ and these concentrations must be developed.

Through direct numerical experimentation, it was determined that the relation between $F$ and the scaled absorbance $A_F$ is linear. If $F$ is too large, loss of model accuracy may ensue due to the onset of inter-particle interactions. Given that for every relevant pair of log-normal distribution parameters $\mu_L$ and $\sigma_L$ a relation between a unit amplitude of $A_F$, denoted as $A_1$, and its corresponding $F$, denoted as $F_1(\mu_L, \sigma_L)$, is known (**Figure 3d**), the formula for $F(\mu_L, \sigma_L)$ with any given $A_F$, can be expressed in equation (eq. S6):

$$F(\mu_L, \sigma_L) = \frac{A_F}{A_1} F_1 = A_F F_1(\mu_L, \sigma_L) \qquad \text{(eq. S6)}$$

Equation (S6) enables the use of $E$ spectra with an arbitrary magnitude, not just that which has a unit amplitude. "ColloidNet" returns the value of $F$ with the probability density of radius, hence equation (eq. S6) is more conceptual rather than practical. The relation between $F$ and concentration of mass ($C_M$) is simply

$$C_M = F\rho \qquad \text{(eq. S7)}$$

where $\rho_{Ag}$ is the density of the NPs, in this work taken as the bulk density of Ag (10.49 g/cm³). Keeping in mind that NPs in the colloid are distributed with a relative abundance $n$ (probability density) based on their radius $R$, the total number of NPs in the volume of optical measurement can be computed:

$$N = \frac{FLr^2}{\frac{4}{3}\sum_i n_i R_i^3} \qquad \text{(eq. S8)}$$

where $L$ is the thickness of the colloid (the interior thickness of the cuvette, in our case 1 mm) and $r$ is the radius (in our case 2.5 mm) of the cylindrical "light volume" with thickness $L$. With this, the concentration of NPs can be computed:

$$C_{NP} = \frac{F}{\frac{4}{3}\pi \sum_i n_i R_i^3} \qquad \text{(eq. S9)}$$

According to **Figure S6 a** (the cross – section is of a single NP), $C_{NP}$ directly impacts the LSPR amplitude. On the other hand, according to[25,26], the interband $E(\lambda)$ is related to the total amount of Ag. This is revealed in **Figure S7a** – the LSPR $E(\lambda)$ magnitude is decreasing with increasing mean NP radius, while the interband $E(\lambda)$ magnitude is increasing. The decrease of $C_{NP}$ for each subsequent growth step occurs due to dilution of the seed solution. It is an inherent part of the seeded growth method and is necessary for the growth of NPs, as shown in[27]. This trend is also observed in the literature describing the similar seeded-growth methodology as used to originate the colloids in this work (**Section S7b**).

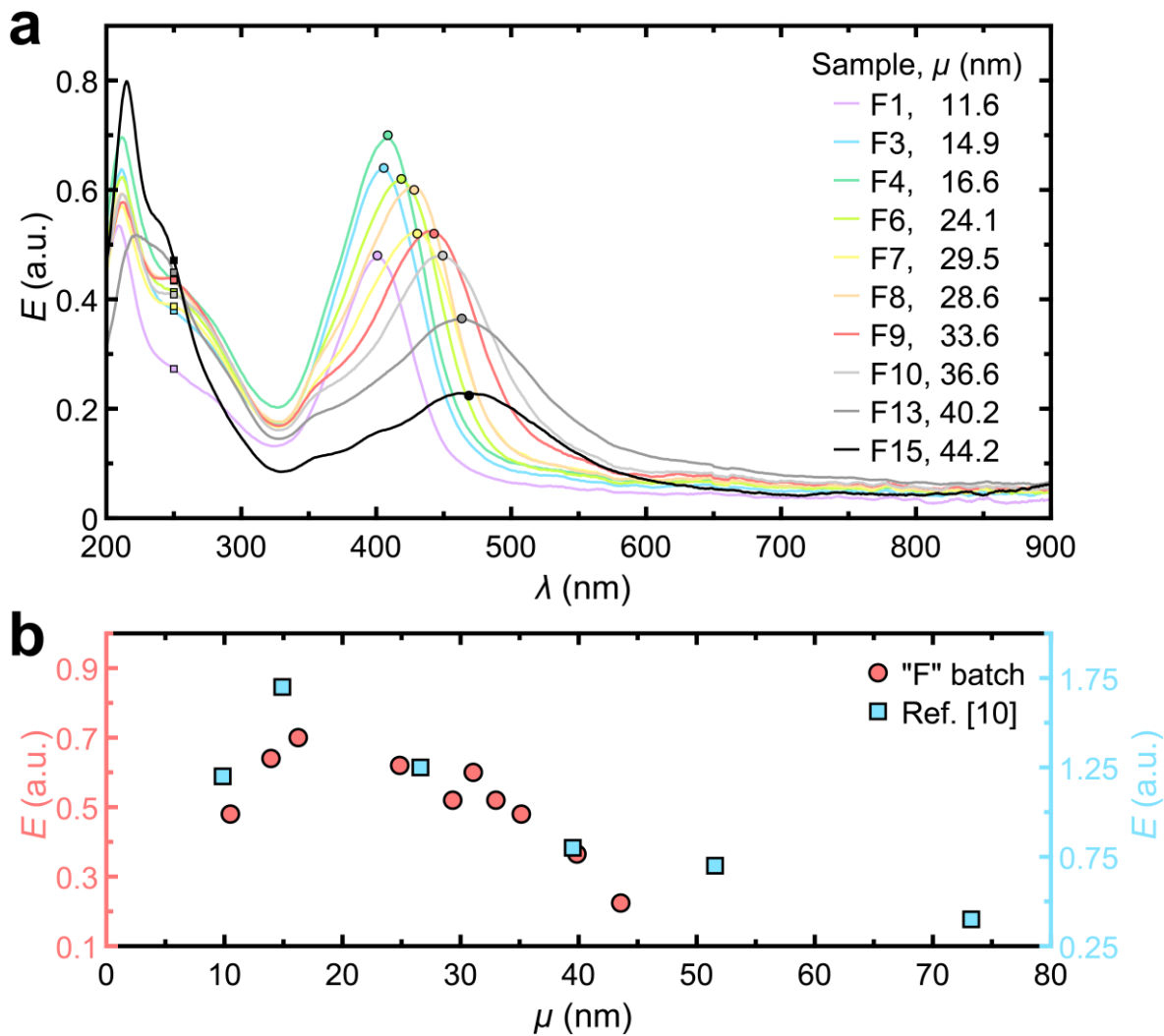

**Figure S7.** $E(\lambda)$ trends for selected colloids of the "F" batch. (a) experimental $E(\lambda)$ spectra of Ag NPs indicating the interband transition at 250 nm and the LSPR peak. (b) The trend of LSPR peak $E(\lambda)$ of the same colloids as in (a, circles) and reported by N. G. Bastús *et al.* in[1] (squares) describing the similar NP synthesis approach.

For monodisperse NPs, it is possible to compute the expected size knowing the concentration of precursor particles (seeds), the mass of NP material added and its density. According to[27], the radius of NPs after a single growth step will be:

$$R_{NP}^3 = R_{seed}^3 + \frac{3}{4} \frac{m_{Ag}}{\pi \rho_{Ag} C_{seed}} \qquad \text{(eq. S10)}$$

where $R_{seed}$ is the radius of the seeds, $R_{NP}$ is the NP radius to which the seeds have grown, $m_{Ag}$ is the added mass of Ag, and $C_{seed}$ is the concentration of seeds. (Eq. S10) assumes no secondary nucleation, making the concentration of NPs equal to that of seeds, and total use of the added material. In order to model NP growth with reduced efficiency of material use, the added mass can be multiplied by a coefficient with a value in the range from 0 to 1.

## S2.6 Additional results of Ag NP size prediction

More examples of predictions by "ColloidNet" can be found in **Figure S8** while **Figure S9** depicts mean sizes of NPs computed based on predictions made by "ColloidNet". In **Figure S8** it can be observed that for NPs with larger mean sizes the prediction of individual histogram bins is poorer. On the other hand, the mean NP size, computed based on the predicted distributions, corresponds better with the ground truth (**Figure S9**). In part, this could be due to the limited resolution of prediction, as radius values are predicted in 1 nm intervals, thus contributing to larger values of RMSPE for samples with a smaller mean size.

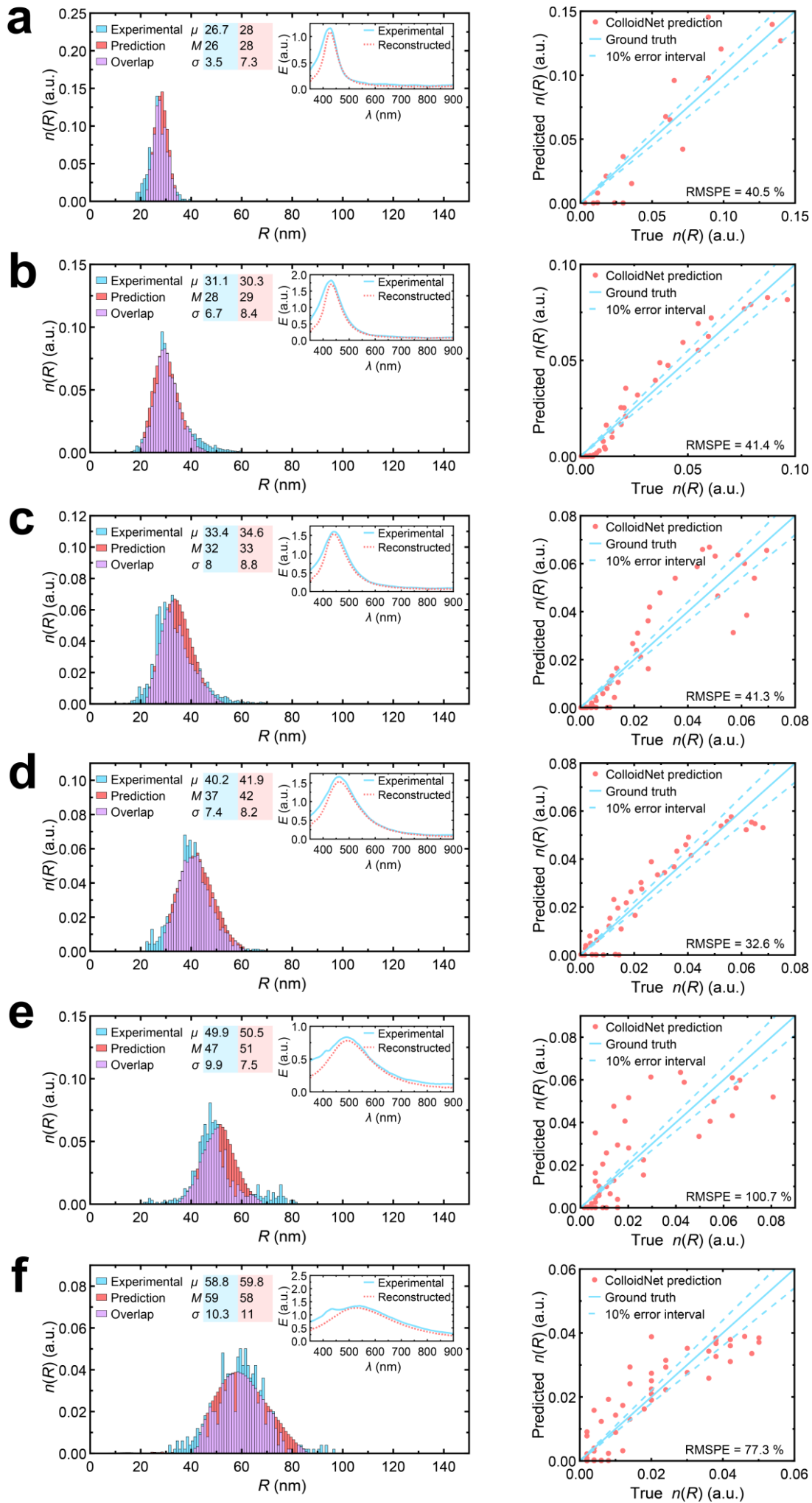

**Figure S8.** Experimental and predicted comparison of full "C" series (sample "C1" (a), "C2" (b), "C3" (c), "C4" (d), "C5" (e), "C6" (f)) NP size distribution characterization by SEM and "ColloidNet. The inset depicts experimental $E(\lambda)$ spectra and dipole spectra reconstructed from the predicted NP distributions and their volume-filling fractions. Next to each case true-predicted value plots for distributions with the estimated RMSPE values and 10% absolute error intervals (dashed lines) are provided.

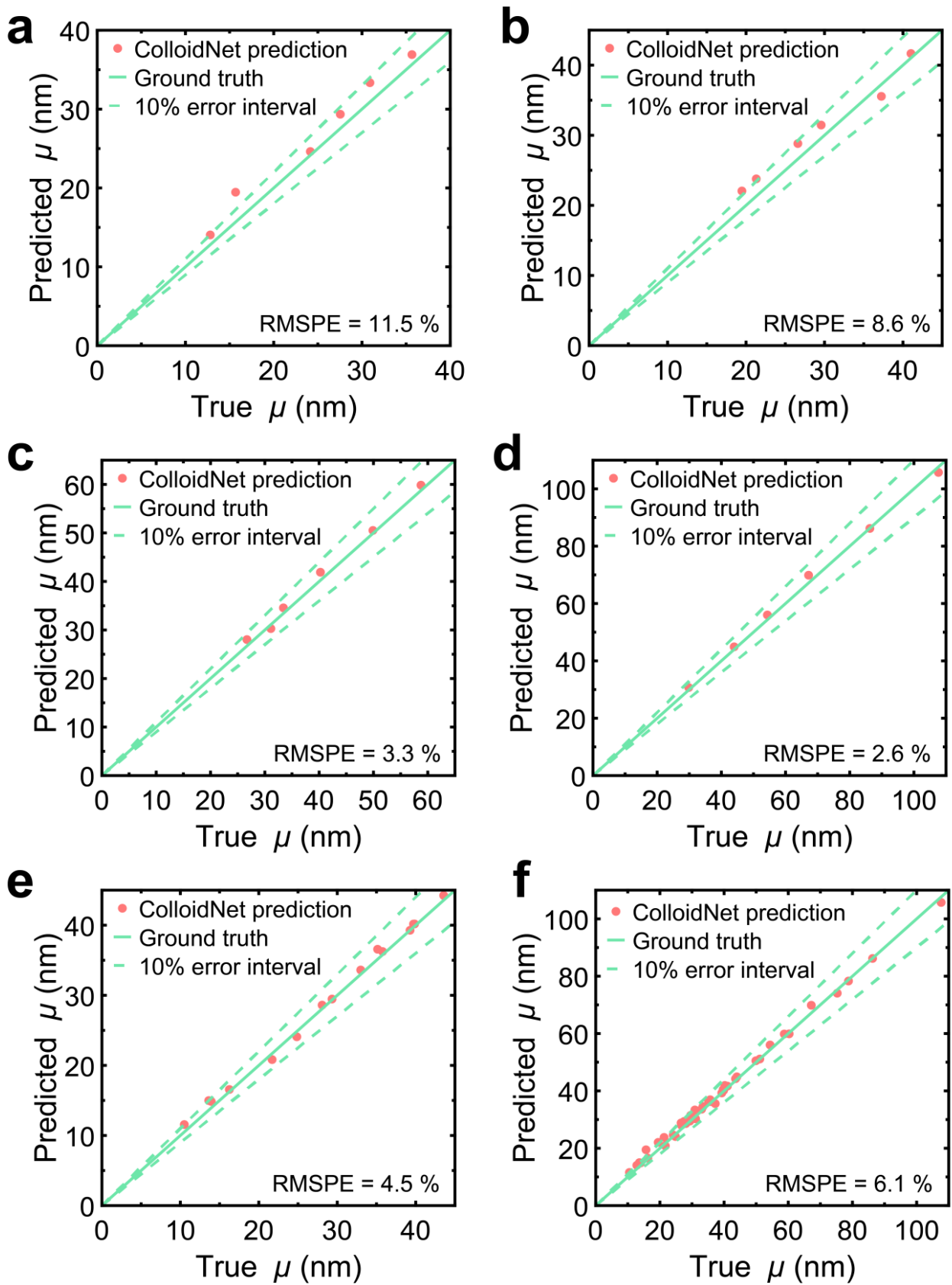

**Figure S9.** Comparison of mean Ag NP size computed based on „ColloidNet" predictions. (a) Batch „A". (b) „B", (c) „C", (d) „D", (e) „F", (f) All batches. Solid lines correspond to the true value while dashed lines correspond to a 10% absolute error interval.